\documentclass[twocolumn,tighten,times]{aastex61}

\received{}
\revised{}
\accepted{}
\submitjournal{ApJ}

\usepackage{graphicx}	
\usepackage{amsmath}	
\usepackage{amssymb}	
\usepackage{color}

\newcommand{\unit}[2]{\ensuremath{\textrm{#1}^{#2}}}

\definecolor{ccolor}{RGB}{127,0,0}

\defcitealias{m16}{M16}
\defcitealias{aa17}{AA17}
\newcommand{\mer}{\citetalias{m16}}
\newcommand{\fhalo}{$f_{\mathrm{halo}}$}

\newcommand{\dform}{\ensuremath{d_{\mathrm{form}}}}
\newcommand{\dpresent}{\ensuremath{d_{\mathrm{present}}}}

\shorttitle{Stellar halo masses}
\shortauthors{Sanderson et al.}

\begin{document}

\title{Reconciling observed and simulated stellar halo masses}

\correspondingauthor{Robyn E. Sanderson}
\email{robyn@caltech.edu}

\author{Robyn E. Sanderson}
\affil{Department of Physics \& Astronomy, University of Pennsylvania, 209 S 33rd St., Philadelphia, PA 19104, USA}
\affil{TAPIR, MC 350-17, California Institute of Technology, Pasadena, CA 91125, USA}\altaffiliation{NSF Astronomy \& Astrophysics Postdoctoral Fellow}
\affil{Columbia University Department of Astronomy, 550 W 120th St, Mail Code 5246, New York, NY, 10027, USA}

\author{Shea Garrison-Kimmel}
\altaffiliation{Einstein Fellow}
\affil{TAPIR, MC 350-17, California Institute of Technology, Pasadena, CA 91125, USA}

\author{Andrew Wetzel}
\affil{TAPIR, MC 350-17, California Institute of Technology, Pasadena, CA 91125, USA}
\affil{Department of Physics, University of California, Davis, CA 95616, USA}
\affil{The Observatories of the Carnegie Institution for Science, Pasadena, CA 91101, USA}

\author{Tsang Keung Chan}
\affil{Department of Physics, Center for Astrophysics and Space Sciences, University of California at San Diego, 9500 Gilman Drive, La Jolla, CA 92093, USA}

\author{Philip F. Hopkins}
\affil{TAPIR, MC 350-17, California Institute of Technology, Pasadena, CA 91125, USA}

\author{Du\v san Kere\v s}
\affil{Department of Physics, Center for Astrophysics and Space Sciences, University of California at San Diego, 9500 Gilman Drive, La Jolla, CA 92093, USA}

\author{Ivanna Escala}
\affil{Department of Astronomy, California Institute of Technology, Pasadena, CA 91125, USA}

\author{Claude-Andr\'e Faucher-Gigu\`ere}
\affil{Center for Interdisciplinary Exploration and Research in Astrophysics (CIERA) and Department of Physics and Astronomy, Northwestern University, 2145 Sheridan Road, Evanston, IL 60208, USA}

\author{Xiangcheng Ma}
\affil{TAPIR, MC 350-17, California Institute of Technology, Pasadena, CA 91125, USA}

\begin{abstract}
We use cosmological hydrodynamical simulations of Milky-Way-mass galaxies from the FIRE project to evaluate various strategies for estimating the mass of a galaxy's stellar halo from deep, integrated-light images. We find good agreement with integrated-light observations if we mimic observational methods to measure the mass of the stellar halo by selecting regions of an image via projected radius relative to the disk scale length or by their surface density in stellar mass . However, these observational methods systematically underestimate the \textit{accreted} stellar component, defined in our (and most) simulations as the mass of stars formed outside of the host galaxy, by up to a factor of ten, since the accreted component is centrally concentrated and therefore substantially obscured by the galactic disk. Furthermore, these observational methods introduce spurious dependencies of the estimated accreted stellar component on the stellar mass and size of galaxies that can obscure the trends in accreted stellar mass predicted by cosmological simulations, since we find that in our simulations the size and shape of the central galaxy is not strongly correlated with the assembly history of the accreted stellar halo. This effect persists whether galaxies are viewed edge-on or face-on. We show that metallicity or color information may provide a way to more cleanly delineate in observations the regions dominated by accreted stars. Absent additional data, we caution that estimates of the mass of the accreted stellar component from single-band images alone should be taken as lower limits.
\end{abstract}




\section{Introduction}
\label{sec:intro}

Recent advances in observational astronomy have begun to reveal the faint stellar halos surrounding Milky-Way-mass galaxies in integrated light \citep{md10,2012arXiv1204.3082B,ds14,m16}. Cosmological simulations \citep{bj05,2007ApJ...666...20P,2008MNRAS.391...14D,font08,cooper10,font11,2013MNRAS.432.3391T,cooper13,pillepich14,2014MNRAS.439.3128T} have long predicted that a significant fraction of the light in these halos, especially at large distances from the main galaxy, should come from the remains of smaller galaxies that were accreted onto the main galaxy and tidally disrupted. Throughout this work we will refer to this component, which is a reflection of the hierarchical nature of structure formation in a cold dark matter (CDM) universe, as the ``accreted stellar component.'' In the CDM picture, the process of hierarchical accretion thus ties the variation in the mass fraction of accreted stars to the accretion history of the host galaxy \citep{bj05,2012MNRAS.420..255T,2013MNRAS.432.3391T,2016MNRAS.458.2371R,amorisco17}. In principle, this component holds some of the few memories of the accretion process. Identifying regions of galaxies that are primarily made up of accreted material is thus the first step toward testing these predictions.

To verify the predictions of CDM, there have been recent attempts to compare the amount of stellar mass observed in the outskirts of galaxies with the mass fraction in accreted stars predicted by simulations \citep[e.g.][]{font08,pillepich14,m16,ds17,2018MNRAS.479.4004E,2018MNRAS.475.3348H}. For purposes of this paper, we will use the term ``stellar halo'' in an observational sense, referring to the faint structure in the outskirts of galaxies beyond the central concentration of stellar mass. In practice there are many observational definitions of this term that can include radial profile, surface brightness, or metallicity characteristics. Most recent observational attempts to characterize the stellar halos of galaxies, including the Milky Way \citep[MW; e.g.][]{2010ApJ...712..692C} and the Andromeda galaxy \citep[M31; e.g.][]{2011ApJ...739...20C}, have used analysis of resolved stellar populations to identify the accreted component, usually by searching for an old, metal-poor population extending far from the central galaxy \citep[e.g.][]{2007IAUS..241..523S,2013MNRAS.428.1248C}. However, this method requires extremely deep images, mainly obtained using the Hubble Space Telescope  \citep[with the exception of][]{2014A&A...562A..73G}, and has therefore been limited to a small handful of galaxies so far. Observing stellar halos in integrated light is in principle more easily scalable to the sample sizes needed to explore the wide variation in the accreted component that is predicted by simulations \citep[e.g.][]{2012arXiv1204.3082B,ds14,2015MNRAS.446..120D,m16,2018MNRAS.475.3348H}, presuming that it is possible to account for the contribution of scattered light \citep{2008MNRAS.388.1521D,2009PASP..121.1267S,2014A&A...567A..97S}. However, the lack of resolved stellar-population information makes it far more challenging to identify the regions of an image dominated by accreted material. So far no work has attempted to account for how the method used to select the stellar halo from a galaxy observed in integrated light may bias the comparison to simulations, where the provenance of material is perfectly known and a variety of definitions of ``stellar halo'' are imposed. Despite efforts such as that in \citet{2016MNRAS.458.2371R} to understand whether the mass in the stellar halo comes primarily from accreted material or from stars formed in the central galaxy (which in this work we call ``formed in situ'') and expelled to the halo, and its dependence with separation from the central galaxy, it is not straightforward to apply these results to the spatial selections in projection that are commonly used in integrated-light images. In fact, most prior work has focused on comparisons between observed galaxies and predictions for the stellar halo based on dark-matter-only (DM-only) simulations tagged with stars, where the stellar halo is by definition 100 percent accreted. However, in simulations that include baryonic physics, both of these distinct channels are observed to contribute to the stellar halos of galaxies \citep{font11,2013MNRAS.432.3391T,cooper13,pillepich15,aa17,gomez17b}, and both are interesting for what they tell us about the process of galaxy formation as well as the cosmology in which galaxies are formed.

\begin{deluxetable*}{lrccDDDc} 

\tablecaption{Simulations used in this work.\label{tbl:sims}}

\tablehead{
 	Name & 
	$m_p\ (M_{\odot})$ &
	$M_{\mathrm{vir}} \ (10^{12} M_{\odot})$ &
	$r_{\mathrm{vir}}$ (kpc) &
	\multicolumn2c{$r_{-2}$ (kpc)} &
	\multicolumn2c{$M_{*,90}\ (10^{10} M_{\odot})$} &
	\multicolumn2c{$r_{*,90}$ (kpc)} &
	Ref 
}

\decimals

\startdata
\textsf{m11f}  & 17000 & 0.50 & 207.7 & 8.5 & 2.5 & 8.3 & C \\
\textsf{m11g}  & 17000 & 0.64 & 225.3 & 8.1 & 4.8 & 8.1 & B \\
\textsf{m12b}  & 56500 & 1.37 & 290.6 & 5.4 & 14.3 & 10.0 & F \\
\textsf{m12c}  & 56500 & 1.30 & 285.9 & 4.7 & 8.9 & 5.5 & F \\
\textsf{m12q}  & 56500 & 1.71 & 313.3 & 3.2 & 16.7 & 6.3 & F \\
\textsf{m12z} & 33000 & 0.87 & 249.9 & 10.2 & 4.0 & 13.0 & H  \\
\textsf{m12m}  & 7070 & 1.47 & 297.8 & 10.7 & 12.6 & 15.7 & L \\
\textsf{m12f}  & 7070 & 1.40 & 293.2 & 14.1 & 8.4 & 16.1 & T \\
\textsf{m12i}  & 7070 & 1.07 & 268.0 & 12.3 & 6.4 & 10.5 & T \\
\textsf{Romeo}  & 28000 & 1.29 & 285.7 & 8.5 & 7.9 & 18.6 & E \\
\textsf{Juliet}  & 28000 & 1.06 & 268.0 & 8.5 & 6.0 & 15.2 & E \\
\textsf{Romulus}  & 31900 & 1.95 & 325.5 & 9.8 & 15.5 & 13.6 & E \\
\textsf{Remus}  & 31900 & 1.25 & 280.4 & 5.1 & 11.5 & 9.0 & E \\
\textsf{Thelma}  & 31900 & 1.44 & 294.4 & 10.7 & 13.2 & 12.8 & E \\
\textsf{Louise}  & 31900 & 1.10 & 269.4 & 5.4 & 7.4 & 14.1 & E \\
\textsf{Batman}  & 57000 & 1.90 & 325.2 & 4.9 & 12.6 & 4.3 & E \\
\textsf{Robin}  & 57000 & 1.58 & 305.9 & 8.5 & 7.1 & 12.2 & E \\	
\enddata

\tablecomments{ 
$m_p$: baryonic particle mass. 
$M_{\mathrm{vir}},r_{\mathrm{vir}}$: \citet{bn98} virial quantities. 
$r_{-2}$: radius where log-slope of dark matter density profile is $-2$.
$M_{*,90},r_{*,90}$: Mass and radius of 90 percent of stellar mass within 30 kpc of the central galaxy at $z\! = \! 0$.} 
\tablerefs{(C) \citet{2017arXiv171104788C}. (B) \citet{2018MNRAS.473.1930E}. (F) Part of the FIRE-2 suite \citep{hopkins17}. (H) \citet{hafen17}. (L) Part of the Latte simulation series \citep{wetzel16}. (E) Part of the ELVIS simulation series \citep[][Garrison-Kimmel et al in prep]{sheagk}.  (T) This work.}

\end{deluxetable*}

Recently, \citet[][hereafter M16]{m16} presented a sample of eight MW-mass galaxies with stellar halos observed in integrated light, from which they estimated stellar halo masses and mass fractions and compared them to predictions from cosmological simulations. Although still small, this is the first such sample to exist in the literature at this mass scale (\citealt{ds14} stacks many galaxies together, and \citealt{2015MNRAS.446..120D} looks at more massive early-type galaxies) and is a promising step towards placing the MW's stellar halo into a cosmological context. Interestingly, most of their measured stellar halo mass fractions lie systematically lower when compared to simulations by nearly an order of magnitude, but both the simulated predictions and the observational methods could potentially have systematic offsets in mass. On the simulation side, selections in present-day, three-dimensional radial distance from the host galaxy are often used to define the halo component, sometimes scaled to the size of the central galaxy in the case of simulations that include baryons \citep[e.g.][]{pillepich14,pillepich15} or with a fixed radial range in the case of tagged DM-only (and hence accretion-only) simulations \citep[e.g.][]{cooper10,cooper13}. On the observational side, \mer\ use a spatial selection in \emph{projected} radius, based on the best-fit disk scale length for the central galaxy, to define the stellar halo component. These differing definitions could be responsible for at least part of the apparent discrepancy between the measured and simulated halo mass fractions reported in \mer. 

In this work we use a set of high-resolution cosmological zoom simulations from the FIRE-2 suite\footnote{See the FIRE project website: \url{http://fire.northwestern.edu}.} \citep{hopkins17} to directly test the consistency and accuracy of the methods used in simulations and observations for measuring the mass fraction in stellar halos of Milky-Way-mass galaxies, and to explore which observational methods might be effective for separating in situ from accreted stars in images. In \S \ref{sec:sims} we describe the simulated galaxies in our sample.  In \S \ref{sec:simhalos} we describe how we define the accreted component of the stellar halo and discuss the general properties of the simulated halos. In \S \ref{sec:mockdata} we describe how we produce mock images from the simulations. In \S \ref{sec:performance} we reproduce methods of measuring the halo mass using our mock data and discuss how the different measurement methods can introduce unwanted biases in halo masses. In \S \ref{sec:improvements} we explore possible ways to reduce these biases, and in \S \ref{sec:conclusion} we summarize.
\newpage

\begin{figure*}
\begin{center}
\includegraphics[width=\textwidth]{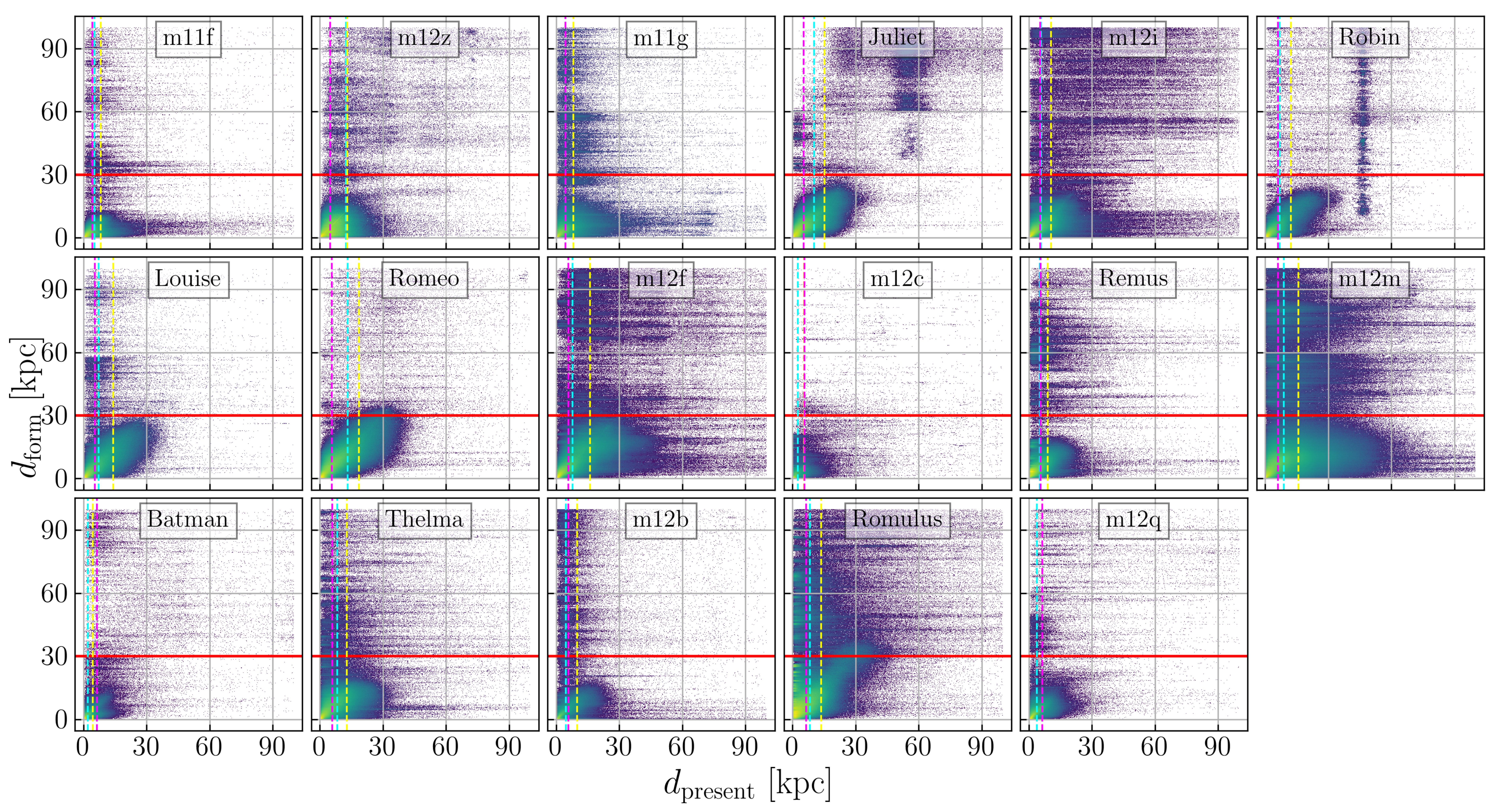}
\caption{Maps of the formation distance \dform\ as a function of present-day distance for star particles within 100kpc of the centers of the simulated galaxies at present day, in order from lowest (top left) to highest (botttom right) stellar mass $M_{90}$. The color scale varies as the base-10 logarithm of the stellar mass density per pixel from low (blue) to high (yellow). Star particles formed beyond 30 kpc from the main galaxy (above the red horizontal line) are considered accreted. Dashed vertical lines indicate $2r_{*,50}$ (cyan, twice the present-day half-mass radius in stars) and $r_{*,90}$ (yellow, radius enclosing 90\% of the stellar mass at present-day, as in Table \ref{tbl:sims}), and $R_{\mathrm{vir}}/50$ (magenta). }\label{fig:dform-vs-dpresent}
\end{center}
\end{figure*}

\section{Simulations}
\label{sec:sims}
In this work we study the stellar halos of a suite of simulated Milky-Way-mass galaxies at different resolutions and with different environments and accretion histories. Their basic properties are summarized in Table \ref{tbl:sims}. All are cosmological zoom-in, hydrodynamical N-body simulations carried out with the GIZMO meshless hydrodynamic simulation code \citep{hopkins15} and FIRE-2 model for star formation and stellar feedback \citep{hopkins17}. In some cases an explicit implementation of subgrid turbulent metal diffusion \citep{hopkins16} is included, which reduces artificial numerical noise in the metallicity distributions \citep{2017arXiv171006533E} but has almost no effect on the large-scale properties of the simulated galaxy \citep{2017MNRAS.471..144S}. For this study we use the highest-resolution simulation available for each set of initial conditions; however, for many of these simulations we also have variations at lower resolution. Appendix \ref{appx:numerics} illustrates that we do not expect these differences to significantly affect the results. 

The suite includes three basic groups of simulations: 
\begin{enumerate}
\item Six isolated halos with resolution of  $57,000\ M_{\odot}$ or better per star particle, with a variety of different accretion histories and stellar masses;
\item Four pairs of halos, with similar resolution to the six isolated halos ($32,000-57,000\ M_{\odot}$ per star particle), selected to roughly resemble the MW-M31 configuration, for a total of eight halos in a Local Group-like environment.  Two of these pairs are hydrodynamical re-simulations of pairs in the ELVIS suite \citep{sheagk}, while the other two are new additions (see Garrison-Kimmel et al. in prep for more details);
\item Three isolated halos simulated at the standard FIRE-2 resolution for $M_h(z=0)\sim10^{12}\ M_{\odot}$ halos ($7070 M_{\odot}$ per star particle). These simulations are part of the ``Latte'' suite first described in \citep{wetzel16}. Two of the three (\textsf{m12i} and \textsf{m12f} in Table \ref{tbl:sims}) have been resimulated to include subgrid turbulent metal diffusion.
\end{enumerate}

The simulations we consider are selected from large cosmological boxes containing thousands of dark matter halos in the mass range $10^{11}$--$10^{12}$ $M_{\odot}$ (see \citealt{2014MNRAS.445..581H} and \citealt{agora} for details of how \textsf{m11} and \textsf{m12} halos are chosen, and \citealt{sheagk} for a description of how the paired halos were chosen). Although the main halos were chosen to fall in specific mass ranges, and in the case of the ELVIS suite to be in pairs with a separation and orbit similar to the MW-M31 system, in all cases these selections were agnostic to the formation histories of the halos \citep{sheagk,wetzel16,2018MNRAS.473.1930E}. \citet{sheagk} further found (in DM-only simulations) no statistical difference between the subhalo properties (counts and kinematics) and host halo properties (formation times and concentrations) of the paired halos in ELVIS compared to similar isolated halos, suggesting that selecting paired halos does not itself result in a selection on the distribution of satellites that form the accreted stellar halo. Our sample, although too small to span the full range, can therefore be considered an unbiased and representative sampling of accretion histories for host galaxies in this mass range. 

FIRE-2 galaxies have already been shown to match the observed stellar mass--dark matter halo mass relation over cosmic time \citep{hopkins17}. This is critical to present-day comparisons of the stellar halo mass fraction in order to ensure that the simulated galaxies falling in this mass range at $z\!=\!0$ have had accretion histories consistent with expectations for this mass scale. The lowest resolution simulations within the group used for this work may slightly overestimate the stellar mass in the main galaxy (see Appendix \ref{appx:numerics}). The FIRE-2 galaxies also feature systems of satellite galaxies at $z\!=\!0$ that are consistent with the mass, size, and number distribution of present-day satellites around the MW and M31 \citep{wetzel16}; although these are not necessarily identical to the building blocks of the stellar halo, this agreement supports the assumption that previously accreted galaxies also had realistic properties and that the disruption rate of satellites by tides is plausible. Finally, the main galaxies in the FIRE-2 suite at this mass scale also match the distribution of observed galaxy \emph{sizes} \citep[][Garrison-Kimmel et al. in prep]{2017MNRAS.467.2430M} which is crucial given that the observational method for selecting the stellar halo involves a spatial cut proportional to the disk scale length.

\section{The stellar halos of simulated galaxies}
\label{sec:simhalos}
\subsection{Distinguishing accreted from in situ stars}
In our simulated galaxies, we distinguish star particles that were accreted from those formed in situ using the distance of each star particle from the center of the main progenitor halo of the $z\! =\! 0$ galaxy of interest, in the first snapshot after it is formed (\dform for short), as in \citet{2017ApJ...845..101B}. The time interval between snapshots is approximately 24 Myr over most of the star formation history, so we consider this to be a suitable approximation to a star particle's ``birth distance.'' This diagnostic is far simpler, but also far less computationally intensive, than the detailed particle-tracking analysis applied to the FIRE-1 simulations by \citet[][hereafter AA17]{aa17}. In terms of their hierarchy of definitions (see Figure 1 of \citetalias{aa17}), this criterion selects stars formed outside the galaxy, whether from externally-processed or non-externally-processed material, that are then incorporated into the galaxy either by merger or intergalactic transfer. Our selection also includes star particles that may have been gas particles at first accretion but form into stars within the accreting galaxy before being tidally stripped, which \citetalias{aa17} would formally consider in situ star formation from externally processed material. Although it fails to capture the nuances in the origin of the star particles, we consider that our formation-distance-based definition adequately captures the accreted stellar component of each simulated galaxy for the purpose of this work, which is to broadly investigate what is being selected using observationally-focused defintions of ``stellar halo.''

As with any such investigation there is necessarily some dependence on the specifics of our criterion for which stars are accreted. To better understand this dependence we consider Figure \ref{fig:dform-vs-dpresent}, which shows the distribution of \dform\ relative to the present-day distance (\dpresent; the distance of that star particle to the galaxy center at $z\! =\! 0$) for all star particles in each of the simulated galaxies out to a present-day galactocentric radius of 100 kpc. In each galaxy the in situ population is visible as a concentration in the lower left-hand corner of the plot that has a roughly 1:1 relationship with large scatter as a function of radius, indicating that that the majority of stars in this population tend to stay near where they are born but may have migrated somewhat \citep[especially the older stars; see][]{2017MNRAS.467.2430M}. Material near either the $x$ or $y$ axis in this diagram is also interesting: along the $y$ axis there is often a fairly significant amount of stellar mass that was formed at large distances from the main halo but is now located in the inner galaxy (though not necessarily in the disk plane), while along the $x$ axis there is material that was formed relatively close in and then scattered out of the inner galaxy. Some tidal streams from disrupted satellites are evident in this diagram as horizontal streaks (that is, they formed as a bound object at some well-defined distance, perhaps in a burst of star formation, and their stars now span a range of radii). Star-forming satellite galaxies are evident as vertical streaks: they are forming new stars at a variety of radii while changing their distance from the main galaxy, but retain all their stars as a bound object with a small present-day distance range. We do not exclude star particles within satellite galaxies from our estimate of the accreted stellar mass, and in fact there are very few galaxies in the inner 50 kpc that have not been tidally disrupted. 

Based on the view in Figure \ref{fig:dform-vs-dpresent} we will define $\dform > 30$ kpc as the separation between in situ and accreted populations. This cut is sufficient to exclude the in situ population for all the galaxies in our sample. We emphasize that this is a conservative choice meant to ensure that very little in situ material is included rather than attempting to capture all the accreted material: clearly some galaxies in Figure \ref{fig:dform-vs-dpresent} have material with  $d_{\mathrm{form}}<30$ kpc that looks like it was accreted. We did check that changing the criterion to $\dform > 20$ kpc does not appreciably change the results. This can be at least partially attributed to the fact that about half of the material designated as ``accreted'' and ending up with \dpresent<100 kpc has \dform>100 kpc, and so does not show up on this figure at all (although its mass is indeed counted as part of the accreted component).  This choice also excludes the small amount of material that formed in the disk and is scattered to large radii by the present day (such as the horizontal streaks near the $y$ axis in, for example, \textsf{m12f}).  We therefore expect that our quoted accreted stellar masses are probably underestimated; Figure \ref{fig:dform-vs-dpresent} gives an idea of the degree of under-estimation for each halo.

Figure \ref{fig:dform-vs-dpresent} also shows, for reference, several length scales commonly used to describe the extent of simulated galaxies or define the region considered the stellar halo in simulations: $2r_{*,50}$ (twice the three-dimensional, spherical radius enclosing 50 percent of the stellar mass within 30 kpc of the central galaxy, shown in cyan; Elias et al. in prep), $r_{*,90}$ (the radius enclosing 90 percent of the stellar mass within 30 kpc of the central galaxy, in yellow; Garrison-Kimmel et al. in prep), and $r_{\mathrm{vir}}/50$, the fraction of the virial radius used in \citet{pillepich14} to define the stellar haloes of Illustris galaxies (shown in magenta).  For most of our galaxies, all three of these length scales are fairly similar and even overlap in some cases, but in most galaxies the in situ sequence (where $\dform \sim d_{\mathrm{present}}$) extends further than any of these distances. For our simulated sample at least, using any of these criteria to select the accreted stellar halo component would over-estimate the mass fairly significantly in most cases. Using a criterion related to a fraction of the stellar mass would also set a maximum value on the stellar halo mass fraction. We therefore avoid using any of these length scales to define the stellar halo.

\subsection{Variation of simulated stellar halo properties}

\begin{figure*}
\begin{center}
\begin{tabular}{cc}
\includegraphics[width=0.5\textwidth]{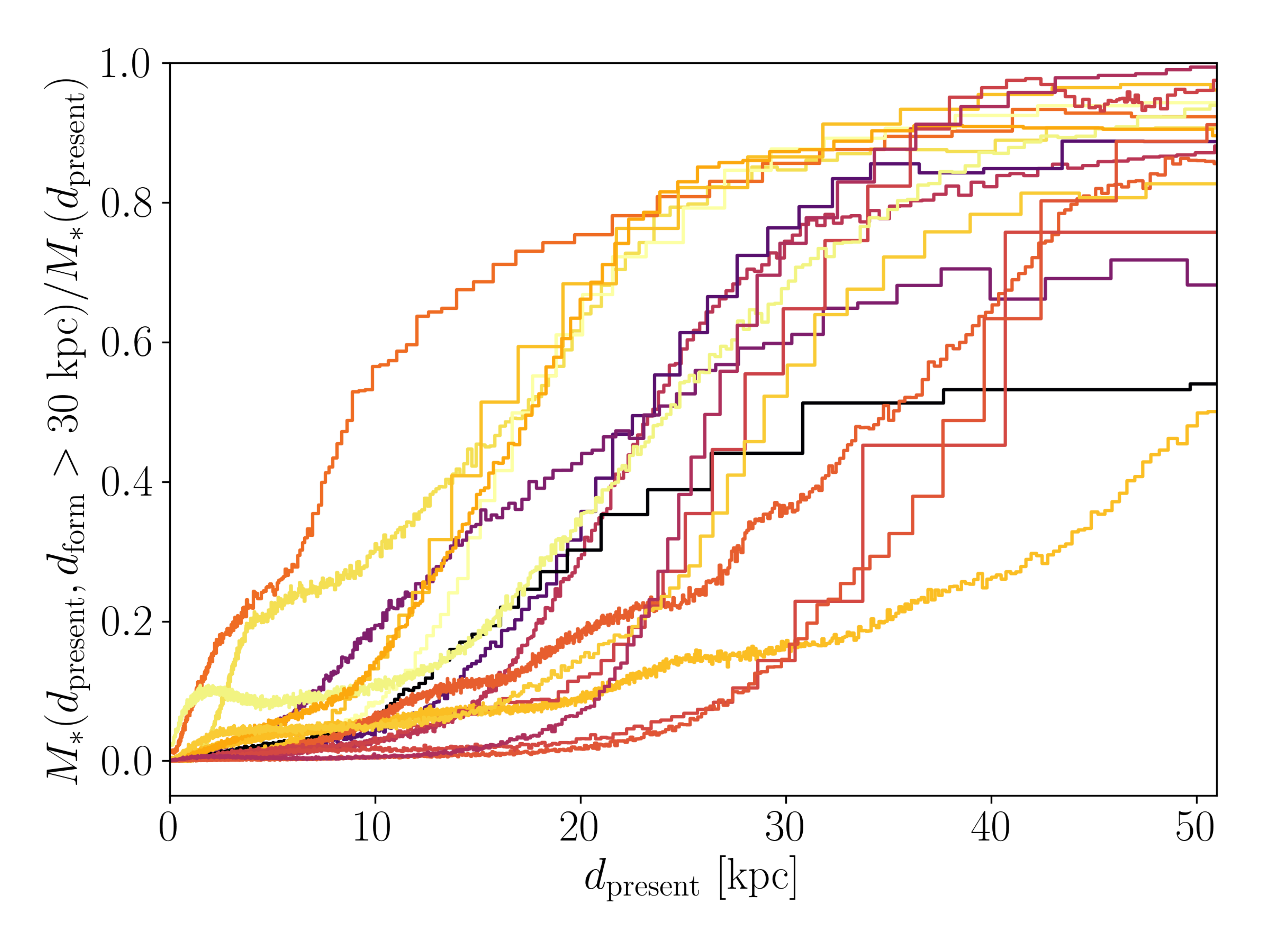} & 
\includegraphics[width=0.5\textwidth]{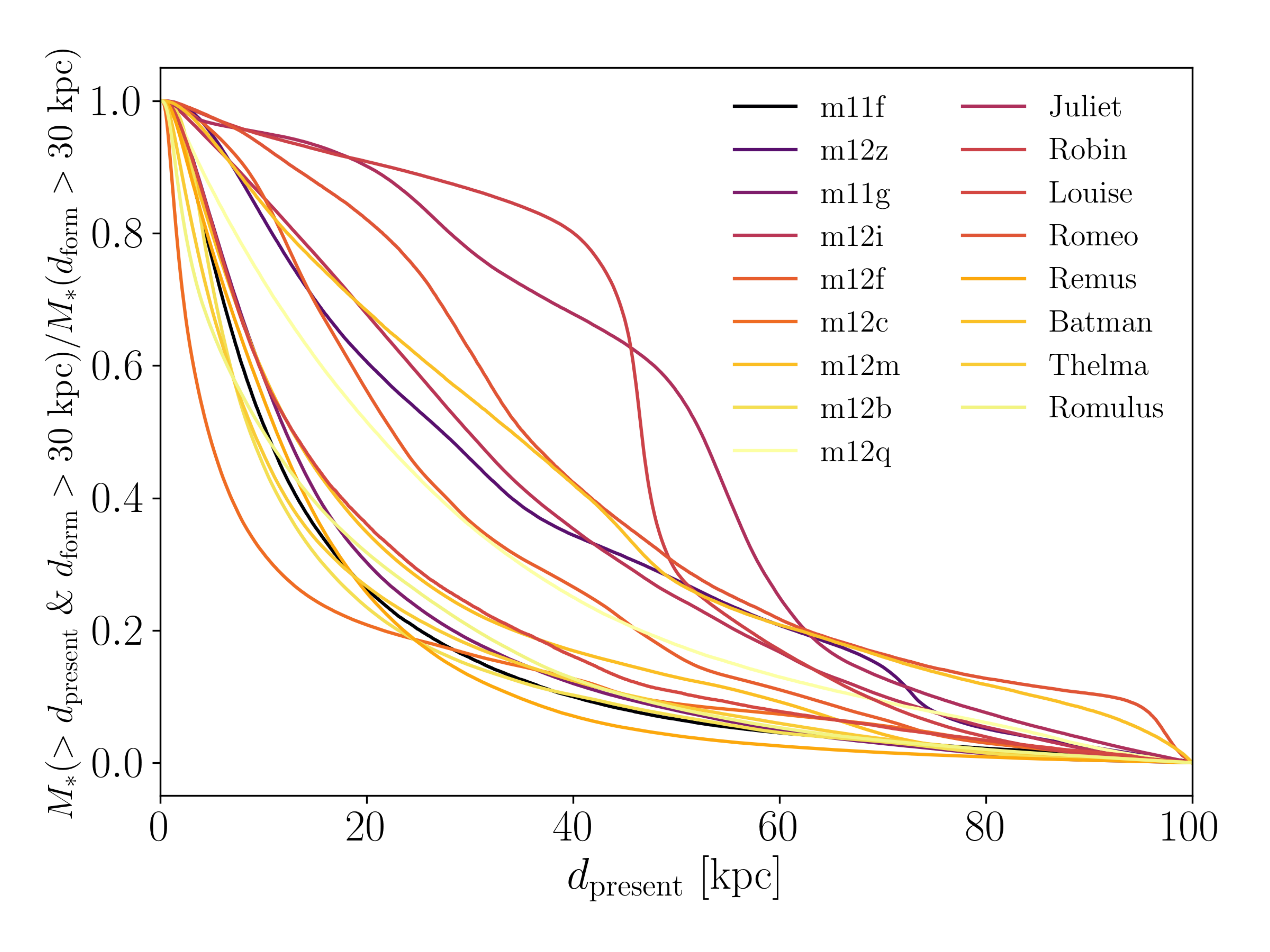}
\end{tabular}
\end{center}
\caption{The transition from in situ to accreted material occurs at a wide range of radii for our simulated galaxies. In both panels, lines are colored by the present-day total stellar mass of the galaxy ($M_{90}$ in Table \ref{tbl:sims}) from lowest (dark purple) to highest (yellow). Left: Fraction of star particles that were accreted (with $\dform>30 kpc$) per bin of 5000 star particles in present-day distance $d_{\mathrm{present}}$, for each simulated galaxy. At 25 kpc from the center, the fraction of material that is accreted ranges from a few to 80 percent. Right: Fraction of stellar mass outside a given present-day distance of the galaxy's center that was accreted (i.e. formed beyond 30 kpc), relative to the total accreted stellar mass presently within 100 kpc of the central galaxy.}
\label{fig:mhalo-cdf}
\end{figure*}

\begin{figure*}
\begin{center}
\includegraphics[width=\textwidth]{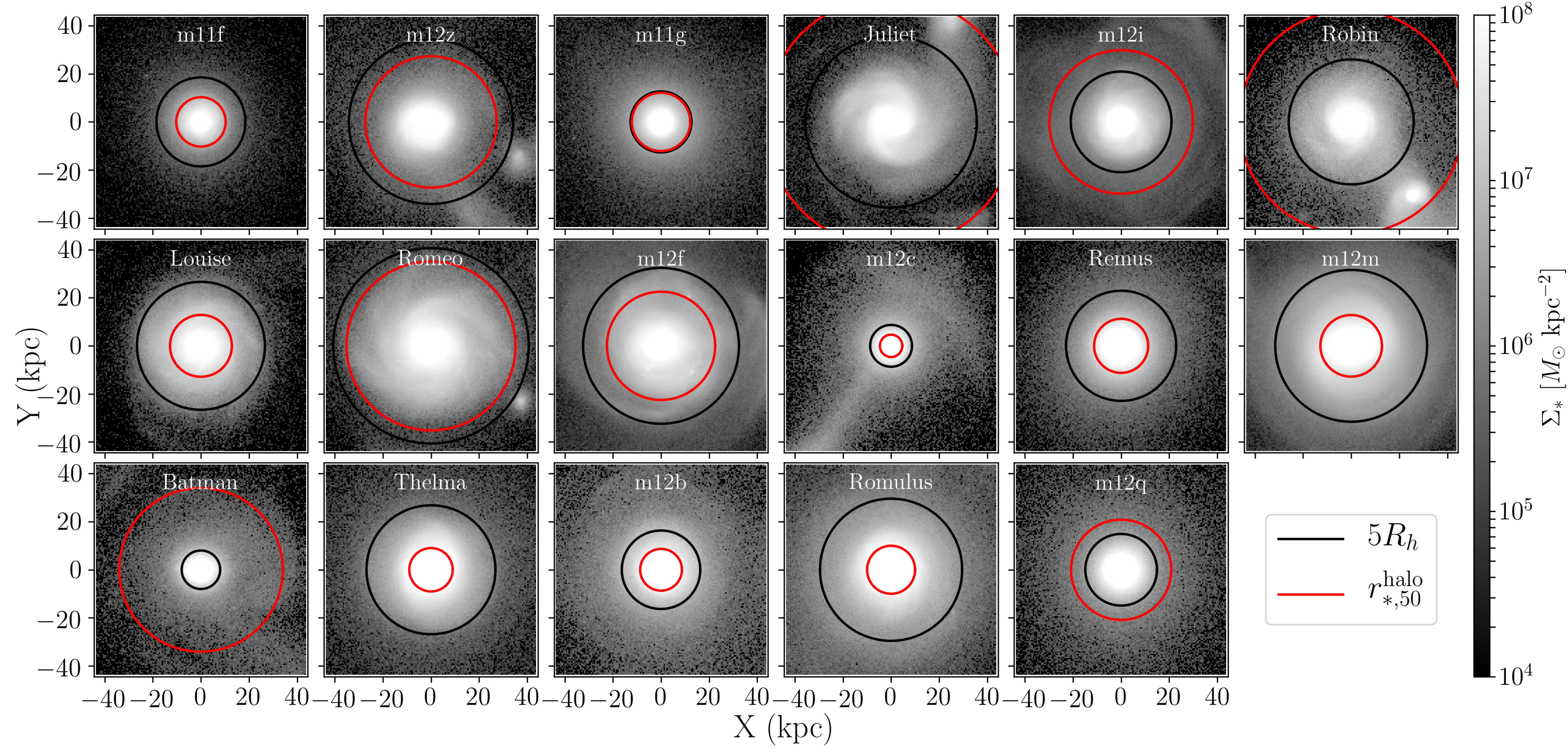}
\includegraphics[width=\textwidth]{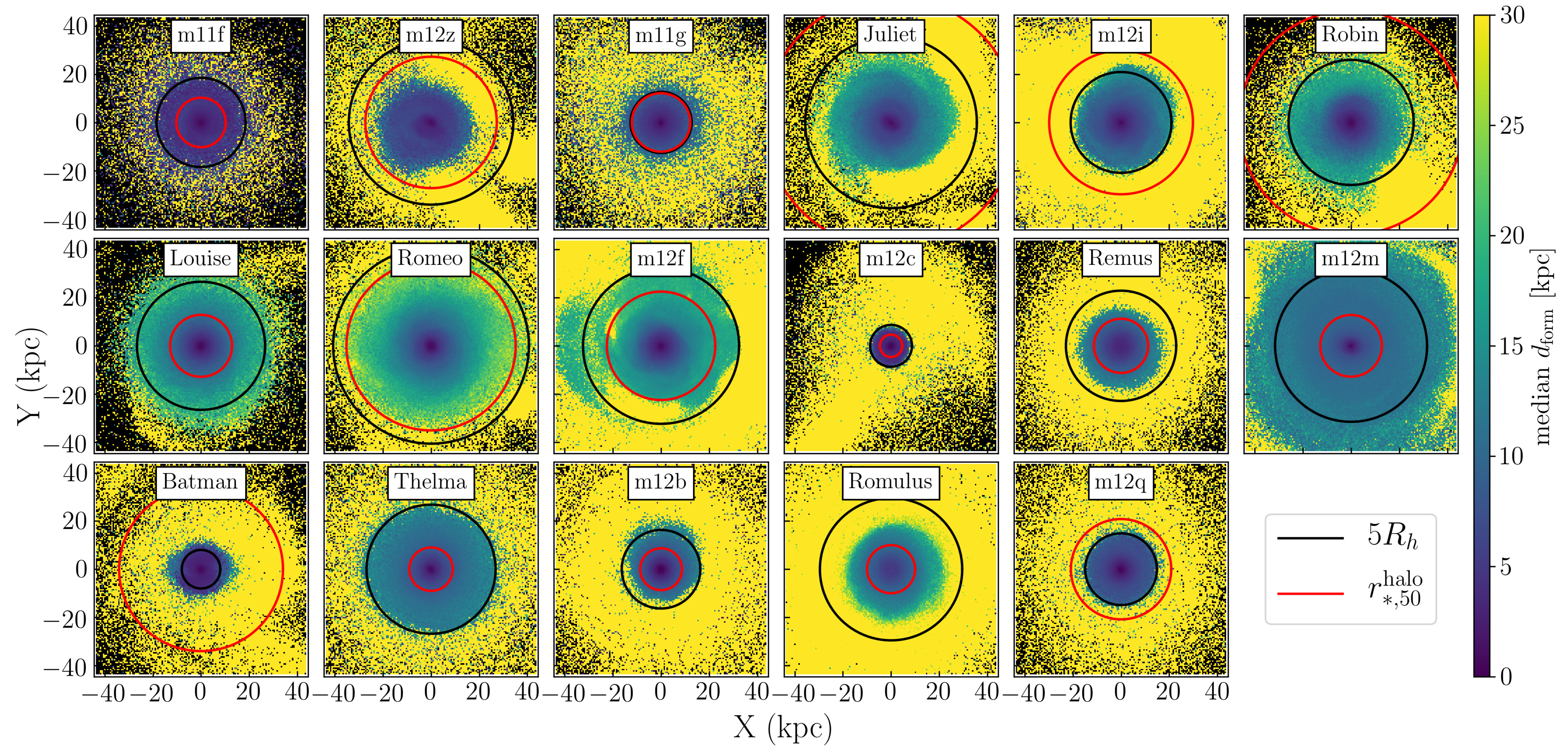}
\caption{Galaxies in our simulated sample show distinct transitions from primarily in situ to primarily accreted stars that are not always accompanied by a distinct transition in the surface mass density.  Top: Simulated stellar mass surface density maps of all the galaxies in our sample, created as described in \S \ref{sec:mockdata}, in order from lowest to highest stellar mass $M_{*,90}$ (Table \ref{tbl:sims}). The log-normalized grayscale shows surface densities between $10^4$ (black; roughly one particle in a pixel in most simulations) and $10^8\ M_{\odot}$ \unit{kpc}{-2} (white) to emphasize fainter outer features. The graininess at the lowest surface densities in the outskirts of each map is due to fluctuations in the number of particles per pixel; no additional observational noise sources were simulated. The pixels used here and in the bottom panels are 12 arcsec on a side, corresponding to 0.6 kpc at the fiducial distance of 10 Mpc used to create these maps. For comparison, central surface densities tend to fall in the range $10^9$--$10^{10}\ M_{\odot}$ \unit{kpc}{-2}, as shown in Table \ref{tbl:fit-results}.
Bottom: Median mass-weighted formation distance of star particles in each pixel of the simulated images in the top panel. Yellow pixels have average $\dform>30$ kpc, our cutoff for material considered accreted (see \S \ref{sec:sims}). Black pixels contain no star particles. The degree to which the region inside $5R_d$ (black circle, see \S \ref{sec:fits}) corresponds to in situ material varies widely from galaxy to galaxy, and does not systematically correlate with the half-mass radius of the accreted stellar halo within 100 kpc, $r^{\mathrm{halo}}_{*,50}$ (red circle; see right panel of Figure \ref{fig:mhalo-cdf}).}
\label{fig:face-on-images}
\end{center}
\end{figure*}

The galaxies in our sample have a wide variety of accretion histories, and differ substantially in the present-day morphologies of both their stellar halos and their disk-bulge systems (Garrison-Kimmel et al. in prep). Figure \ref{fig:mhalo-cdf} shows how the wide variation in the accretion histories of the galaxies is reflected in the radial distribution of their accreted material. The left-hand panel shows a broad diversity in the fraction of accreted material, relative to the total stellar mass within 50 kpc, as a function of radius (computed using bins of 5000 star particles). The range of radii where the transition from mostly in situ to mostly accreted material occurs is quite broad, consistent with the results of \citet{2016MNRAS.458.2371R} for a much larger sample of simulated galaxies in this stellar mass range. In the right-hand panel of the figure, the $y$ axis shows the fraction of accreted material (defined as $\dform\! >\! 30$ kpc) presently at distances larger than $x$, relative to the total accreted mass within 100 kpc, for each simulated galaxy. The degree of central concentration of the accreted material also varies widely from halo to halo: the half-mass radius of the accreted component can be anywhere from 5 to 50 kpc, varying by an order of magnitude for a range of about a factor 7 in host stellar mass. For comparison, the ratio of largest to smallest $r_{90}$ within our sample, which measures the extent of the \emph{total} stellar mass, is 4.3. No trend with host stellar mass is apparent in Figure \ref{fig:mhalo-cdf}. An investigation of the Spearman rank coefficient, which was used to test for correlations between the half-mass radius of the accreted material and the host stellar mass, bears this out: the value of $r_\mathrm{Sp}=-0.24$ is in the 16th percentile of correlation values computed for a bootstrapped sample of 1000 shuffled versions of the same data. These are approximately normally distributed, thus the computed value is roughly $1\sigma$ from the median of the decorrelated samples, indicating a statistically insignificant degree of correlation.

Examining the outliers in Figure \ref{fig:mhalo-cdf}, we find that some of them would be readily apparent from images of the galaxy while others are less so. In cases where the accreted component is unusually extended (like \textsf{Robin}, \textsf{Romeo}, and \textsf{Juliet}) a relatively massive companion contributes a substantial fraction of accreted material at relatively large separation, while at the opposite extreme (like \textsf{m12c}) nearly all the accreted material is within 20 kpc of the galactic center at the present day. Tellingly, these two extremes are in different stages of a relatively major interaction whose signatures are still present in images of the galaxy (see Figure \ref{fig:face-on-images}). Even excluding such cases, however, one is left with the difference between galaxies like \textsf{m12f} and \textsf{m12b}, neither of which show visible signs of a recent merger with a large companion but which have substantial differences in the concentration of their accreted material: only 20 percent of \textsf{m12b}'s accreted stars lie beyond 20 kpc, while 70 percent of  \textsf{m12f}'s do. \textsf{m12f} did in fact have a recent merger with a galaxy of roughly the original mass of the Sagittarius dwarf, that contributes to its large and extended halo and has substantially disrupted its outer disk, but this is not immediately apparent from the mock image alone.

The face-on projections shown in the top panel of Figure \ref{fig:face-on-images} show how this maps onto the present-day appearance of the different simulated galaxies. The sample considered in \mer\ tends toward spiral galaxies with extended disks (like \textsf{m12i}, \textsf{m12f}) but also includes a few with prominent bulges (like \textsf{m12b} and \textsf{Batman}). 
 
The masses and mass fractions of the stellar halos in our suite are broadly similar to the general trends found in studies that used semi-analytic modeling \citep{bj05} or particle tagging in DM-only simulations \citep{cooper10,cooper13}, or in larger cosmological volumes simulated at lower resolution using different physical models than those in FIRE \citep{font11,2013MNRAS.432.3391T,pillepich14,pillepich15,2016MNRAS.458.2371R}.  Because (as we will show in \S \ref{sec:sim-estimates}) the strategies used to define halo mass make an important difference, we will not make more detailed comparisons between our results and those of simulations by other groups using different codes. Instead, we will focus in the rest of this work on something that the high-resolution galaxies here can do uniquely well: testing the fidelity of observational methods for measuring the mass in accreted stars.

\section{Mock observations of simulated stellar halos}
\label{sec:mockdata}
In order to compare different observational methods for determining stellar halo mass, we first need to create a mock ``observation'' of each galaxy to which we can apply the method. Given that the goal of this paper is not to explore the validity of assumptions about color corrections, stellar population modeling, or correction for background light, we chose to produce and compare maps of stellar mass surface density, rather than modeling the spectral energy distribution (SED) to produce mock images in different bands. This approach necessarily assumes that the process of converting from flux in a set of filters produces an accurate estimate of the stellar mass surface density in each pixel.

We attempt to reproduce as closely as possible the method used in \mer\ to determine \fhalo\ for our simulated galaxies, which comprises the following steps:
\begin{enumerate}
\item Obtain a map of the stellar mass surface density for each system. \mer\ uses a color transformation on images; we bin the star particles in the simulation as described in \S \ref{sec:maps}.
\item Fit a double S\'ersic profile to the disk and bulge (see \S \ref{sec:fits}).
\item Sum the stellar mass in pixels with ellipsoidal radius \emph{larger} than 5 times the scale length of the disk component in the S\'ersic fit to obtain the mass in the stellar halo.
\item Divide by the \emph{total} stellar mass in the image to obtain the stellar halo mass fraction, \fhalo.
\end{enumerate}

\subsection{Maps of stellar mass surface density}
\label{sec:maps}
We place each simulated galaxy at a distance of 10 Mpc, roughly the median of the \mer\ sample, and rotate it to a face-on orientation defined by calculating the principal axes of the stellar mass distribution within 10 kpc of the Galactic center. We then create 60x60 arcmin maps of stellar mass surface density (corresponding to about 85x85 kpc) made up of pixels 12 arcsec on a side (about 0.6 kpc) by binning the star particles of the simulated galaxy in projection. The mock images are shown as thumbnails in the top panel of Figure \ref{fig:face-on-images}. Consistent with the approach in \mer, we do not expressly attempt to remove satellite galaxies from the mock images.

Spiral arms are faint but evident in the disks of many of our simulated galaxies. In the outskirts, faint structure is seen in the stellar halos that in most cases extends beyond the frame of the map. This is expected, since the maps only probe to about 50 kpc in projected radius while the typical virial radii of the dark matter halos for our simulated galaxies are about 300 kpc (see Table \ref{tbl:sims}).  

\begin{deluxetable*}{lDccDcc}

\tablecaption{Results of profile fitting for simulated galaxies.\label{tbl:fit-results}}

\tablehead{
    \colhead{Name} & 
    \twocolhead{$M_*$ $(10^{10} M_{\odot})$} & 
    \colhead{$I_d$ $(10^9 M_{\odot} \unit{kpc}{-2})$} & 
    \colhead{$R_d$ (kpc)} & 
    \twocolhead{$I_b$ $(10^9 M_{\odot} \unit{kpc}{-2})$} & 
    \colhead{$R_b$ (kpc)} & 
    \colhead{$n$}
}

\decimals
\startdata
\textsf{m11f} & 2.8 & 0.1 & 3.7 & 0.8 & 1.0 & 0.9 \\
\textsf{m11g} & 5.4 & 0.5 & 2.5 & 3.3 & 0.7 & 0.6 \\
\textsf{m12b} & 16.1 & 0.7 & 3.3 & 7.4 & 0.8 & 0.7 \\
\textsf{m12c} & 10.1 & 1.5 & 1.7 & 11.0 & 0.6 & 0.5 \\
\textsf{m12q} & 18.7 & 1.1 & 3.0 & 7.7 & 1.0 & 0.6 \\
\textsf{m12z} & 4.6 & 0.1 & 6.8 & 0.3 & 1.0 & 0.6 \\
\textsf{m12m} & 14.3 & 0.2 & 6.3 & 2.3 & 1.3 & 1.2 \\
\textsf{m12f} & 9.5 & 0.1 & 6.5 & 2.2 & 1.2 & 0.8 \\
\textsf{m12i} & 7.2 & 0.2 & 4.2 & 2.0 & 1.0 & 0.9 \\
\textsf{Romeo} & 9.0 & 0.1 & 8.1 & 0.7 & 1.5 & 0.9 \\
\textsf{Juliet} & 6.9 & 0.1 & 7.1 & 0.6 & 1.6 & 1.2 \\
\textsf{Romulus} & 17.7 & 0.3 & 5.9 & 3.0 & 1.3 & 0.8 \\
\textsf{Remus} & 12.9 & 0.3 & 4.6 & 3.4 & 1.3 & 0.8 \\
\textsf{Thelma} & 14.8 & 0.3 & 5.4 & 3.6 & 1.0 & 0.7 \\
\textsf{Louise} & 8.2 & 0.2 & 5.3 & 2.3 & 1.0 & 0.7 \\
\textsf{Batman} & 14.1 & 2.5 & 1.6 & 23.9 & 0.5 & 0.6 \\
\textsf{Robin} & 8.9 & 0.2 & 5.2 & 1.9 & 1.1 & 0.7 \\
\enddata

\tablecomments{See Equation \eqref{eq:sersic1} for explicit definitions of fit parameters. $M_*$: Total stellar mass in frame of mock image (see \S\ref{sec:mockdata}). $I_d$: Central surface density of disk component. $R_d$: Scale radius of disk component. $I_b$: Central surface density of bulge component. $R_b$: scale radius of bulge component. $n$: S\'ersic index of bulge component. }

\end{deluxetable*}

\subsection{Galaxy profile fitting}
\label{sec:fits}

As in \mer\, we describe the disk and bulge of each simulated galaxy using a double S\'ersic profile as a function of ellipsoidal projected radius $\tilde{R}$: 
\begin{eqnarray}
\label{eq:sersic1}
\Sigma_*(\tilde{R}) &=& I_b  \exp \left\{ -b_n \left[ \left( \frac{\tilde{R}}{R_b} \right)^{1/n} -1 \right] \right\} + \nonumber \\
&&+ I_d  \exp \left\{ -b_1 \left[ \left( \frac{\tilde{R}}{R_d} \right) -1 \right] \right\}
\end{eqnarray}
with $b_n$ defined such that the effective radius ($R_b$ for the bulge or $R_d$ for the disk) contains half the total luminosity. We fit this function to the region of each simulated stellar mass surface density map out to approximately the last spiral features or drop-off in surface density. For most galaxies this region is within 20 to 30 kpc in $\tilde{R}$. We restrict the scale radius $R_b$ of the bulge component to be less than or equal to the scale radius $R_d$ of the disk component, fix the index of the disk component to 1 (i.e. we assume an exponential disk), and allow the index $n$ of the bulge component to vary between 0 and 5. We set a broad prior on the logarithm of the two normalizations $I_b$ and $I_d$ of the bulge and disk components respectively. 

In \mer\ the outermost spiral features in the galaxy images were used to select the region used to fit a S\'ersic profile to the disk and bulge, but since we use stellar mass density directly rather than synthesizing a spectral energy distribution, spiral features are often less apparent than in an image, since young stars are much brighter in luminosity than they are massive. If no clear spiral features are present we use the radius corresponding to a drop-off in surface mass density. For all of our galaxies this is roughly 20--40 kpc; the results of the S\'ersic fit are not sensitive to the exact value.

In order to account for the fact that galaxies are not perfectly round, the mass and light profiles of galaxies are commonly described as a function of ellipsoidal projected radius $\tilde{R}$. Instead of determining the ellipsoidal radius of each pixel by fitting ellipsoids in annuli as was done in M16, we directly compute them from the same principal-axis vectors used to rotate the disks into a face-on projection, such that
\begin{equation}
\tilde{R}^2 \equiv x^2 + \left(\frac{a}{b} y\right)^2,
\end{equation}
where $a$ and $b$ are the axis ratios of the principal axes defining the directions $x$ and $y$, respectively. This choice is not as sensitive to possible twists in the orientation of the disk as a function of projected radius, but since most of our simulated galaxies are fairly regular in shape the resulting scatter in the density profile tends to be fairly small; hence we do not expect it to significantly affect the fitted parameters. We also checked that the by-eye choice of radial range for the fit does not significantly affect the results.

We perform a least-squares fit to the surface mass density of the individual pixels for each galaxy using Levenberg-Marquardt minimization with a fit-by-eye first guess. However, since this algorithm finds only the nearest local minimum, we also did a more complete Monte Carlo sampling of parameter space using \textsf{emcee} \citep{emcee} for one galaxy, to understand how the well-known degeneracies inherent in S\'ersic fitting affected the determination of the disk scale length $R_d$ which is used to delineate between disk and halo in \mer. The details of this test are described in Appendix \ref{appx:fit-test}; we conclude that even when other parameters are degenerate the value of $R_d$ is sufficiently robust that we can use the Levenberg-Marquardt  minimization results (Table \ref{tbl:fit-results}) with confidence.

\begin{figure*}
\begin{center}
\includegraphics[width=\textwidth]{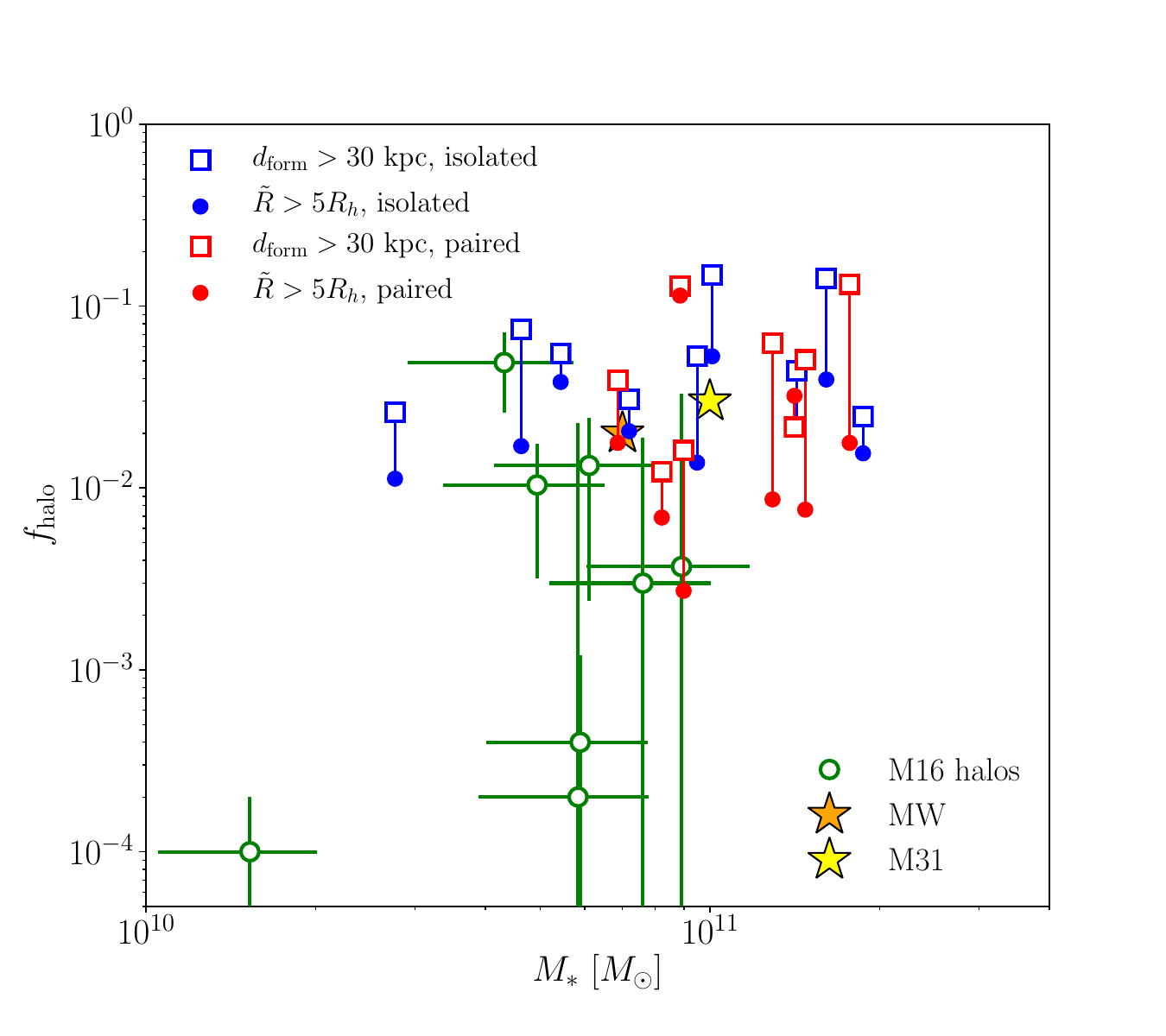}
\caption{Stellar halo mass fractions $f_{\mathrm{halo}}$ calculated for the simulated galaxies in Table \ref{tbl:sims} (red and blue points), compared with the observed galaxies in \mer\ (green). Estimates from resolved stellar populations for the MW (\citealt{2010ApJ...712..692C}, orange) and M31 (\citealt{2011ApJ...739...20C}; yellow) are given for context. Filled red and blue symbols show estimates using fitted values of $R_d$ given in Table \ref{tbl:fit-results} to distinguish accreted from in situ stars (the same method as in \mer); open symbols show the actual accreted mass fraction in our simulations using $d_{\mathrm{form}}>30$ kpc. Pairs of estimates for the same simulated galaxy are connected with vertical lines. Data for the simulations are given in Table \ref{tbl:halo-mass-estimates}. For simulated galaxies with similar total stellar mass to the observed sample, we obtain halo fractions that are statistically consistent with those measured by \mer, with no systematic difference between paired and isolated halos. The true accreted fraction in our simulated galaxies is systematically higher, as seen in previous work.}
\label{fig:halo-fractions}
\end{center}
\end{figure*}

\section{Performance of typical methods to measure stellar halo mass}
\label{sec:performance}
We examine two existing observational methods for measuring the stellar halo mass fraction \fhalo: the disk-fitting method employed by \mer\ and the surface mass density cutoff motivated by the \citet{cooper13} retagging of DM-only halos in the Millennium-II simulation at the low end of their mass range (see their Figure 6). We also look at the performance of three selections used in simulations for separating stellar halo from disk populations, to determine whether the simulated and observed methods agree on stellar halo mass when applied to the same simulated galaxy. All these methods rely on a selection in either present-day spatial location (3D or projected) or surface density (which is generally correlated with position) to define the accreted stellar halo, but we see in Figure \ref{fig:dform-vs-dpresent} that while \dform\ and \dpresent\ are tightly correlated for most stars formed in situ, this is not the case for accreted material. Accreted stars can have large \dform\ but small \dpresent\ due to dynamical friction; adding to the confusion, some material formed in situ can be removed to large \dpresent where a spatial cut on \dpresent\ would mistake it for accreted stars. The accreted material at small \dpresent\ could make up only a small fraction of the total mass at these distances, but a large part of the total \emph{halo} mass. In essence, therefore, we are asking whether any of these spatial selections are successful as a relatively unbiased proxy of the total mass in accreted stars.

\begin{figure*}
\begin{center}
\includegraphics[width=\textwidth]{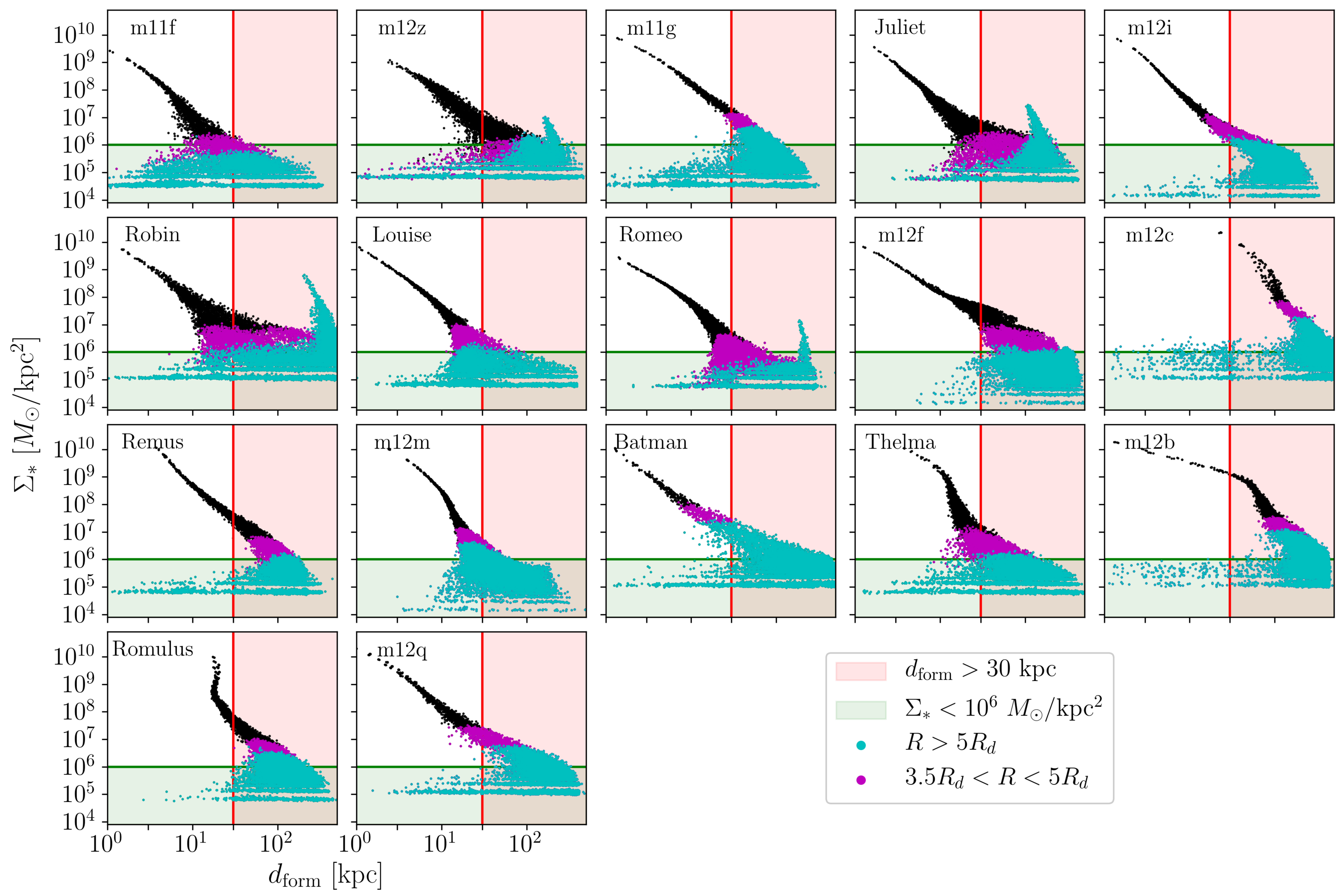}
\caption{Distribution of stellar mass surface densities in individual pixels (black points) as a function of the mass-averaged formation distance \dform\ of star particles in each pixel. Red-shaded region is dominated by ``accreted stellar halo;'' that is, pixels where stars have mass-weighted median $\dform>30$ kpc. Magenta (cyan) points have $\tilde{R}>3.5R_d$ ($\tilde{R}>5R_d$). Green-shaded region shows pixels with $\Sigma_* <  10^{6}\ M_{\odot}$ \unit{kpc}{-2} (see \S \ref{sec:fhalo-sigma}). The typical surface density at the transition from stars formed in situ to accreted material ranges over three orders of magnitude, from $10^6$ to $10^9$ $M_{\odot}$ \unit{kpc}{-2}. The wide variation in the relationship between surface density and stellar origin underlines the difficulty of separating accreted material using a surface-density criterion.}
\label{fig:rform-dists}
\end{center}
\end{figure*}

\subsection{Stellar halo mass fraction estimates from profile fitting}
\label{sec:fhalo-rh}

We calculate \fhalo\ for each of our simulated galaxies in a manner analogous to \mer, by summing the stellar mass in pixels with ellipsoidal radius $\tilde{R}>5R_d$, subtracting the extrapolated mass of the disk and bulge in the fitted profiles, and dividing by the total stellar mass $M_*$ within the field of view. The results of this calculation are shown as filled symbols in Figure \ref{fig:halo-fractions} along with the results from \mer\ and estimates for the MW \citep{2010ApJ...712..692C} and M31 \citep{2011ApJ...739...20C}. To determine the accuracy of this method for estimating \fhalo, we also calculated the fraction of stellar mass in each simulated galaxy, within the field of view of the mass map, that was formed more than 30 kpc from the center of the galaxy. This quantity (i.e., the true accreted fraction) is plotted with open symbols in Figure \ref{fig:halo-fractions}. When we reproduce the observational method on our simulated galaxies, we obtain halo fractions that agree with those measured by \mer, but the \emph{accreted} mass fraction in our simulated galaxies is systematically higher than this.

Compared to the galaxies analyzed by \mer, our simulated galaxies tend to have slightly higher total stellar mass. Remarkably, some of the simulated galaxies have stellar masses and \fhalo\ values quite close to those estimated for the MW (\textsf{m12i} and \textsf{Juliet}) and M31 (\textsf{m12f}); these methods use resolved stellar populations and so in principle should not be subject to the same biases as the integrated-light estimates. The weak trend toward lower \fhalo\ at lower stellar mass apparent in the \mer\ galaxies is also present in our simulations, and appears to connect smoothly with their lower-mass galaxies in terms of ``observed'' stellar halo mass fraction (closed symbols). The large variation in \fhalo\ at a given stellar mass is also apparent in our simulated galaxies, although our sample is too small to assess the full extent of the scatter. Galaxies simulated in pairs resembling the MW and M31 (red points) do not appear to have systematically different \fhalo\ than those evolved in isolation; we expect this is because our mock observations probe a relatively small region around each galaxy (out to 50 kpc) relative to the separation of the pairs (about 1 Mpc).

The lowest mass fraction estimated in \mer, which occurs for the lowest-mass galaxy in the sample, is still significantly below the range of stellar halo mass fractions in our suite of simulations. However, we note that this galaxy (M101) has one of the most extended disks in the sample, which, as we will discuss shortly, corresponds to the most severe under-estimation scenario in our simulated sample. We also expect the scatter in halo fractions to increase significantly at lower total stellar mass \citep{cooper13,ds14,pillepich15}: the building blocks of accreted stellar halos in galaxies of lower mass are themselves lower in mass, and therefore have a larger scatter in stellar mass that may be at least partially responsible for this trend \citep[e.g.][]{cooper10}. Although our sample is unbiased with regard to formation history (see \S \ref{sec:sims}) it is too small to understand whether M101 is consistent with the tails of the distribution; Elias et al. (in prep) explores this question using the Illustris simulations.

It is also evident from Figure \ref{fig:halo-fractions} that the choice of $5R_d$ as the dividing line between in situ and accreted populations underestimates \fhalo\ in nearly every galaxy. The extremely steep fall-off of most stellar halos with distance means that the estimate of the mass in the stellar halo is extremely sensitive to the radius at which this selection is made, and for most of our halos $5R_d$ is well outside the region where most of the stars have {$d_{\mathrm{form}}>30$ kpc}. To illustrate this, in the bottom panel of Figure \ref{fig:face-on-images} we show the average formation distance of the star particles in each pixel of the simulated face-on images shown in the top panel, with a black circle indicating $5R_d$, and a red circle indicating the half-mass radius of the stellar halo within 100 kpc (see the right-hand panel of Figure \ref{fig:mhalo-cdf}), superposed for each galaxy. It is clear that selecting material outside $5R_d$ to represent the halo is a far better assumption for some galaxies than others. It is also clear that while for some galaxies the accreted material is quite centrally concentrated, in others it is much less so, with the half-mass radius well outside $5R_d$. In these cases (such as \textsf{Juliet}) one could indeed hope to identify a region that includes most of the accreted stellar mass while still excluding the in situ material, while observationally disentangling the majority of the stellar halo mass in galaxies like \textsf{m12b} may prove impossible.

\begin{figure*}
\begin{center}
\includegraphics[width=\textwidth]{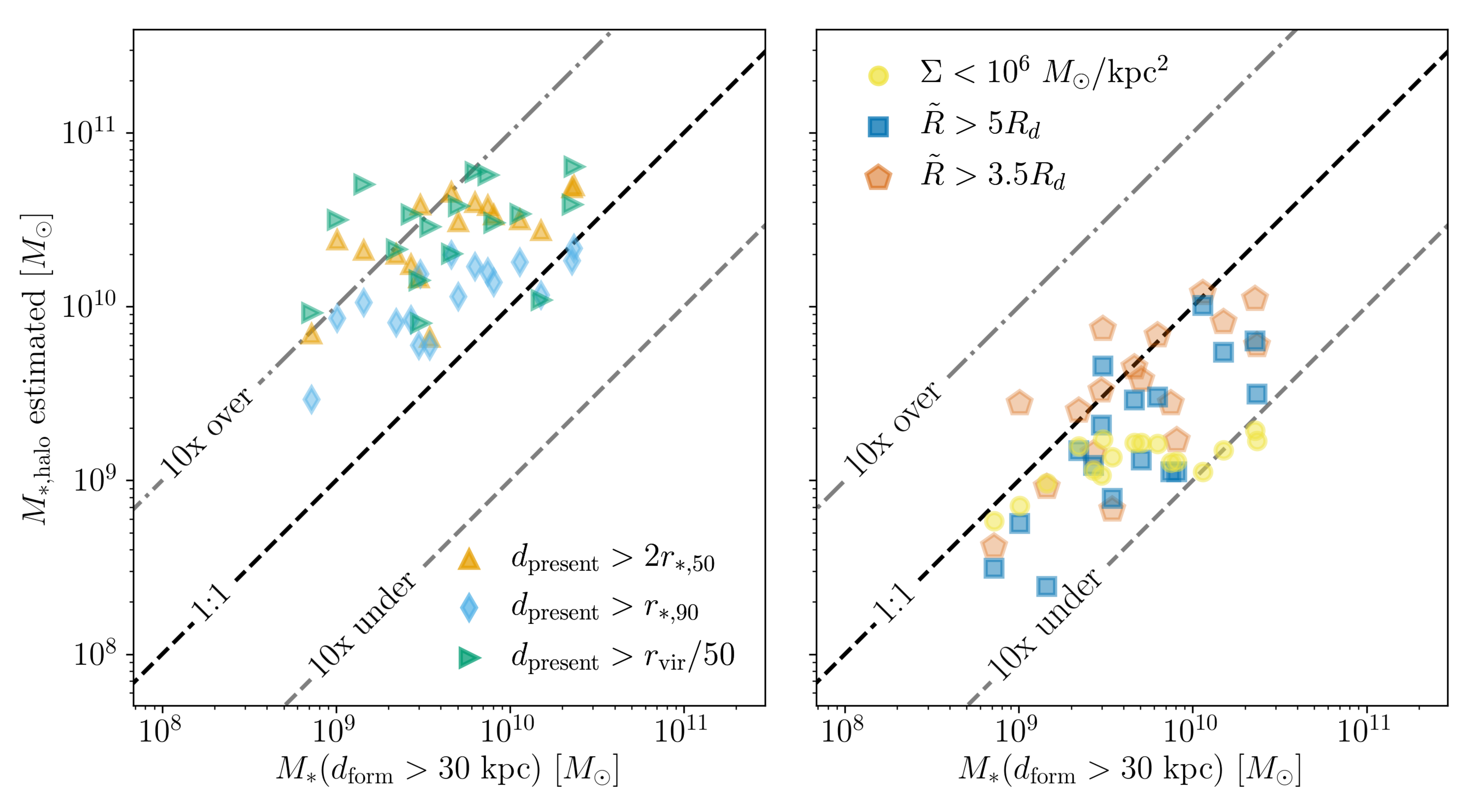}
\caption{Comparison of the extent to which different definitions of ``stellar halo'' correspond to the true accreted stellar mass (stars formed beyond 30 kpc from the main galaxy). Left: criteria commonly used to select stellar halos of simulated galaxies (\S \ref{sec:sim-estimates}). All stars with present-day distance $\dpresent>x$, for each of the listed thresholds $x$, are considered part of the stellar halo; this tends to \emph{over}estimate the mass in accreted stars because the cuts include a significant amount of material formed in situ (see Figure \ref{fig:dform-vs-dpresent}). Right: definitions based on observed galaxy images (\S\S\ref{sec:fhalo-rh}--\ref{sec:fhalo-sigma}), where all stars below a threshold surface density or with projected ellipsoidal radius $\tilde{R}$ (\S \ref{sec:fits}) larger than some value are considered part of the stellar halo. This tends to \emph{under}estimate the accreted stellar mass when applied to our simulated galaxies viewed face-on. Blue squares show the criterion used in \mer, yellow circles the criterion proposed in \citet{cooper13}, and orange pentagons show a recalibrated version of the \mer\ criterion chosen to produce an unbiased estimate for this sample (see extended discussion in the text). The dashed lines in both panels show unbiased estimates (black) and a factor of 10 under- or over-estimate (gray) for reference.}
\label{fig:mass-estimates}
\end{center}
\end{figure*}

Comparing the top and bottom panels of Figure \ref{fig:face-on-images}, it is not clear immediately whether the morphology of the galaxy could be used to diagnose whether selection by projected radius is likely to give a good estimate of the halo-dominated region. Nevertheless, we can ask whether simply using a smaller characteristic radius can give a more consistent estimate of \fhalo. We find that using $\tilde{R}>3.5R_d$ as the cutoff (plotted as orange pentagons in the right-hand panel of Figure \ref{fig:mass-estimates}) gives a relatively unbiased result across the limited mass range of stellar halos in our sample, with over- and underestimates all less than 0.5 dex.

Given the wide variety of distributions of accreted material apparent in Figure \ref{fig:face-on-images}, it is not clear whether a superficial adjustment to the cutoff radius used to select the halo is really a complete solution to the problem. To understand better how using a radial selection may be biasing stellar halo mass estimates, and to investigate further whether a galaxy's morphology could be a clue to calibrate estimates of the stellar halo mass, we looked for correlations between the true or estimated stellar halo mass and three basic parameters: the total stellar mass of the galaxy, the extent of the disk (represented by $R_d$), and the bulge-to-disk mass ratio $M_b/M_d$ (computed by integrating the best-fit S\'ersic profile to 5 effective radii for each component). The \emph{true} accreted halo masses have no apparent dependence on the extent of the stellar disk, but as expected due to the steep falloff in the mass profile of most stellar halos, a larger stellar disk will induce a greater underestimate of the stellar halo mass when $\tilde{R}>5R_d$ is used as the cutoff. Because of the correlation between $R_d$ and $M_*$, the bias from using $\tilde{R}>5R_d$ can artificially suppress the dependence of \fhalo\ on $M_*$. Adopting $\tilde{R}>3.5R_d$ seems to mitigate this issue somewhat. We see no dependence of stellar halo mass on the S\'ersic index of the bulge or the bulge-to-disk ratio.

Figure \ref{fig:rform-dists} plots the distribution of surface densities and average formation distances for all the pixels in the face-on galaxy images. The pixels selected by the wider radial cut $5R_d$ (cyan points) often do not include the pixels with an average formation distance beyond 30 kpc (the red shaded region) at the highest stellar mass surface densities: in other words, any black point falling in the red shaded area represents stellar mass that should be counted as accreted stellar halo but is excluded by the radial selection. Changing to $3.5R_d$ rather than $5R_d$ (magenta points) appears to help by including primarily brighter pixels; in some cases this also agrees better with the selection based on \dform but in others it includes material formed closer to the main galaxy than our criterion considers accreted (although many of the galaxies shown here and in Figure \ref{fig:dform-vs-dpresent} have disks that do not extend all the way to $\dform=30$ kpc). Reassuringly, however, using a smaller disk exclusion region tends to incorporate material that was still formed at appreciably large distances from the main galaxy.

\begin{deluxetable*}{lDDcccDDD}

\tablehead{ 
\colhead{Name}& 
\twocolhead{$M_*$} &
\twocolhead{$5{R}_{d}$  (kpc)} &
\colhead{$M_*(\dform\! >\! 30\ \textrm{kpc})$} & 
 \colhead{${M_*}(\tilde{R}\! >\! {5{R}_{d}})$} & 
 \colhead{${M_*}(\Sigma\! <\! 10^{6})$} &
\twocolhead{$f_{\mathrm{halo}}^{\dform}$}   & 
\twocolhead{${f}_{\mathrm{halo}}^{5{R}_{d}}$}   & 
\twocolhead{ ${f}_{\mathrm{halo}}^{\Sigma}$ }
}

\tablecaption{Estimates of stellar halo mass.\label{tbl:halo-mass-estimates}}

\decimals

\startdata  
\textsf{m11f} & 2.8 & 18.5 & 0.07 & 0.04 & 0.06 & 0.026 & 0.011 & 0.021 \\
\textsf{m11g} & 5.4 & 12.7 & 0.30 & 0.22 & 0.11 & 0.055 & 0.038 & 0.020 \\
\textsf{m12b} & 16.1 & 16.3 & 2.28 & 0.66 & 0.19 & 0.142 & 0.040 & 0.012 \\
\textsf{m12c} & 10.1 & 8.7 & 1.51 & 0.56 & 0.15 & 0.149 & 0.053 & 0.015 \\
\textsf{m12q} & 18.7 & 14.9 & 0.46 & 0.31 & 0.16 & 0.025 & 0.016 & 0.009 \\
\textsf{m12z} & 4.6 & 34.2 & 0.35 & 0.09 & 0.14 & 0.075 & 0.017 & 0.029 \\
\textsf{m12m} & 14.3 & 31.6 & 0.63 & 0.32 & 0.16 & 0.044 & 0.021 & 0.011 \\
\textsf{m12f} & 9.5 & 32.4 & 0.51 & 0.14 & 0.16 & 0.053 & 0.015 & 0.017 \\
\textsf{m12i} & 7.2 & 20.9 & 0.22 & 0.16 & 0.16 & 0.031 & 0.021 & 0.022 \\
\textsf{Romeo} & 9.0 & 40.7 & 0.14 & 0.04 & 0.10 & 0.016 & 0.003 & 0.011 \\
\textsf{Juliet} & 6.9 & 35.6 & 0.27 & 0.13 & 0.11 & 0.039 & 0.018 & 0.017 \\
\textsf{Romulus} & 17.7 & 29.6 & 2.34 & 0.34 & 0.17 & 0.132 & 0.018 & 0.010 \\
\textsf{Remus} & 12.9 & 22.9 & 0.81 & 0.13 & 0.13 & 0.062 & 0.009 & 0.010 \\
\textsf{Thelma} & 14.8 & 26.8 & 0.75 & 0.14 & 0.13 & 0.051 & 0.008 & 0.009 \\
\textsf{Louise} & 8.2 & 26.6 & 0.10 & 0.07 & 0.07 & 0.012 & 0.007 & 0.009 \\
\textsf{Batman} & 14.1 & 8.0 & 0.31 & 0.47 & 0.17 & 0.022 & 0.032 & 0.012 \\
\textsf{Robin} & 8.9 & 26.0 & 1.14 & 1.02 & 0.11 & 0.129 & 0.114 & 0.013 \\
\enddata 

\tablecomments{All masses are in units of $10^{10}\ M_{\odot}$. 
$M_*$: total stellar mass in field of view (see Figure \ref{fig:face-on-images}). 
$5{R}_{d}$: projected distance threshold used in \mer\ to distinguish disk and halo (see \S \ref{sec:fits} and Table \ref{tbl:fit-results}). 
$M_*(\dform\! >\! 30\ \textrm{kpc})$ ($f_{\mathrm{halo}}^{\dform}$): Mass (mass fraction) in stellar halo determined by selecting material with $\dform>30$ kpc (see \S \ref{sec:simhalos}). 
${M_*}(\tilde{R}\! >\! {5{R}_{d}})$ ( ${f}_{\mathrm{halo}}^{5{R}_{d}}$): Mass (mass fraction) in stellar halo determined by selecting pixels with ellipsoidal radius $\tilde{R}>5R_d$ (see \S \ref{sec:fhalo-rh}). 
${M_*}(\Sigma\! <\! 10^{6})$ (${f}_{\mathrm{halo}}^{\Sigma}$): Mass (mass fraction) in stellar halo determined by selecting pixels with $\Sigma_*<10^6\ M_{\odot}$ \unit{kpc}{-2} (see \S \ref{sec:fhalo-sigma}). }

\end{deluxetable*}

For many of the galaxies in our sample there is a significant transition between a relatively well-correlated sequence in formation distance at high surface densities, driven in part by star formation in situ, to a more scattered picture at lower surface density. Material at small \dform\ is not part of the \emph{accreted} stellar halo under our definition, but nonetheless exists at large separation from the main galaxy and low stellar mass surface density in many cases, and could fairly be considered part of the stellar halo by many definitions. However this makes up a very small portion of the stellar halo mass (usually less than a percent) for most of our simulated galaxies. 

Using a less conservative selection of $\tilde{R}>3.5R_d$ (shown as magenta points) seems to result in less bias over the entire sample mainly because this selection appears to do a slightly better job getting most of the pixels with average $\dform>30$ kpc at the cost of occasionally including some material at lower \dform. Given the strong correlation between surface density and \dform, and the relatively conservative nature of this criterion to begin with, most of the overshoot is still material that was formed relatively far from the galaxy, or is at very low surface densities as discussed above. Using a smaller radius also means the differences between fitted profiles are less exaggerated, especially given that some of the visible disks in the mock images of Figure \ref{fig:face-on-images} seem to fade out before $5R_d$, and that the bottom panel of Figure \ref{fig:face-on-images} shows that material that looked like part of an extended disk could actually be considered accreted for many of the galaxies with the largest $R_d$. Practically speaking, one could envision optimzing a selection criterion over a larger sample of simulated galaxies (especially spanning a wider range in total stellar mass), but Figure \ref{fig:rform-dists} underlines how the correspondence between stellar mass surface density and stellar origin differs so widely between galaxies that \emph{any selection based on projected radius should be used with caution}.

\subsection{Estimation methods used in simulations}
\label{sec:sim-estimates}
In simulations of galaxy halos, a spatial cut is sometimes used to try to separate accreted from in situ material, defined in terms of the three-dimensional rather than projected radius. In Figure \ref{fig:dform-vs-dpresent}, which shows three examples of cutoff radii used in other works ($r_{vir}/50$, $2r{_*,50}$, and $r_{*,90}$; see Section \ref{sec:simhalos}), we showed that these choices tend to fall within the outskirts of the disk sequence for many of the FIRE simulated galaxies, which would indicate that these cuts are likely to overestimate the mass in the stellar halos by mistakenly including stars formed in situ. The left panel of Figure \ref{fig:mass-estimates} illustrates that this is indeed the case: all these selection criteria tend to over-estimate the stellar mass in the halo, and the bias is worse for lower-mass halos. These results may change for simulations in which the stellar-to-dark-matter-halo mass ratio (and hence the distribution of $r_{vir}$ at a given stellar mass) or the size distribution of galaxies at a given stellar mass differs substantially from our sample, illustrating the importance of matching the observed distributions of these properties with samples of simulated galaxies. 

\subsection{Stellar halo mass fraction estimates based on surface density criteria}
\label{sec:fhalo-sigma}
We also considered using a surface density criterion to define the stellar halo as an alternative to the profile-based method used in \mer. To compare with simulations, \mer\ defined the simulated tagged stellar halos of the Aquarius simulations \citep{cooper10} as the region where $\Sigma_* < 10^{6}\ M_{\odot}$ \unit{kpc}{-2} (shown as the green shaded region in Figure \ref{fig:rform-dists}). For most galaxies in our sample, this selection picks out the same approximate set of pixels as one of the selections on projected radius, with a few exceptions where it is more conservative. We apply this criterion to our simulated stellar mass surface density maps and compare it to the value obtained by selecting stars with $d_{\mathrm{form}}>30$ kpc, as we did for the $R_d$-based estimate in \S \ref{sec:fhalo-rh}. The yellow circles in the right-hand panel of Figure \ref{fig:mass-estimates} show how this strategy corresponds to the mass in accreted stars.

Although both of the observational methods shown in Figure \ref{fig:mass-estimates} (excluding the criterion specifically calibrated on this data set) systematically underestimate the mass of the stellar halo, the $R_d$-based selection does better than the surface density criterion. Remarkably, although the masses of the stellar halos in our simulated galaxies (as defined by formation distance) range over more than an order of magnitude, the mass in pixels with surface densities less than $10^6\ M_{\odot}$ \unit{kpc}{-2} is nearly constant over this entire range.

\begin{figure*}
\begin{center}
\includegraphics[width=\textwidth]{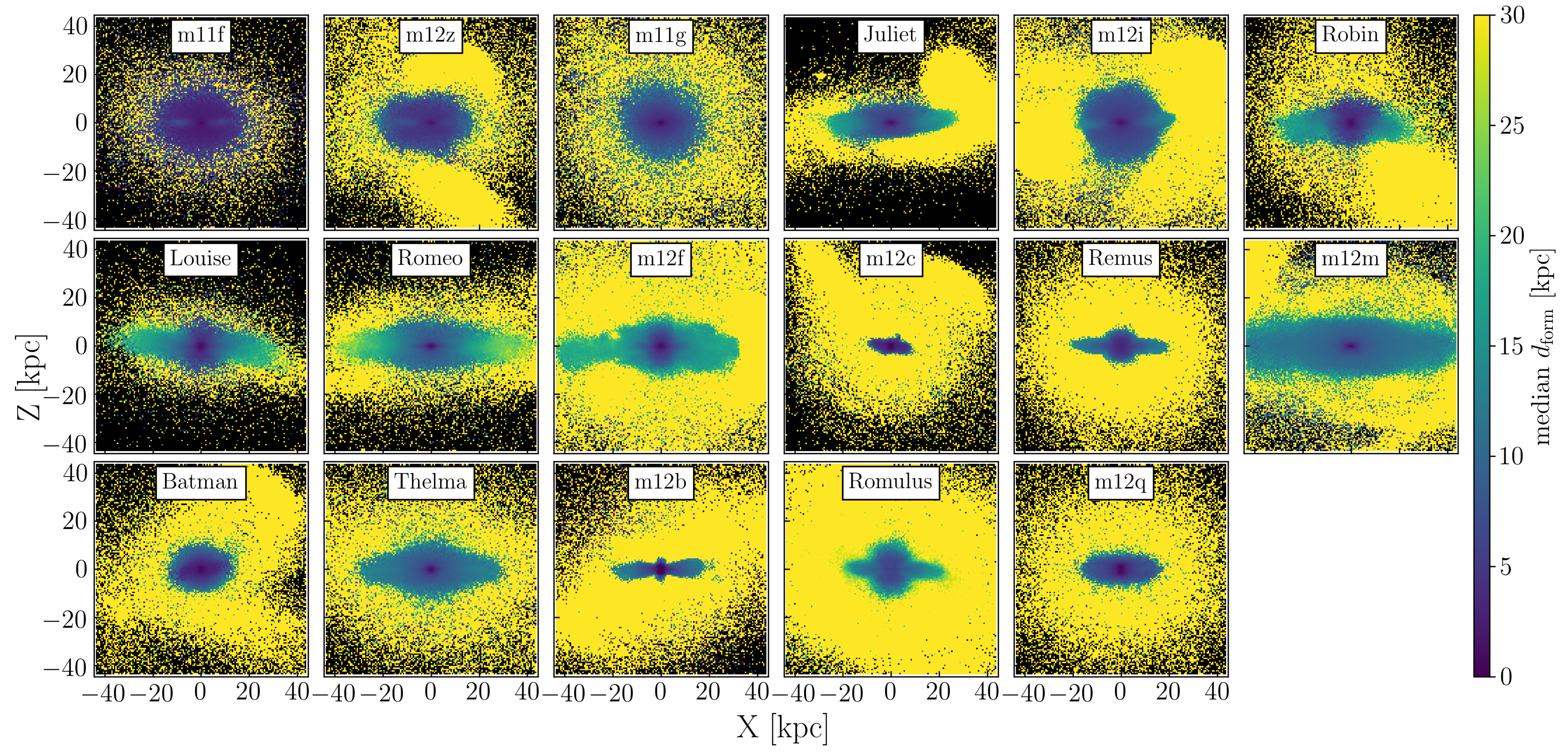}

\caption{Edge-on images of mass-weighted median \dform, as in Figure \ref{fig:face-on-images}. In some galaxies (\textsf{m12i}) an in situ component extends well above the thin disk plane, while in other cases (\textsf{m12b}) in situ material is closely confined to the thin disk plane. In other galaxies (\textsf{Robin}, \textsf{m12f}) interactions with massive satellites warp the in situ component; in later stages of mergers the accreted material dominates the in situ component entirely (\textsf{m12c}). Although the disk generally subtends less area than in a face-on orientation, the great variety of accretion histories still complicates the determination of halo stellar mass in edge-on galaxies.}
\label{fig:edge-on-sucks-too}
\end{center}
\end{figure*}

\section{Improved accreted halo mass fractions}
\label{sec:improvements}
Given that the observational methods we have tried so far tend to \emph{under}estimate the accreted halo mass, while the methods used in simulations tend to \emph{over}estimate it, we now consider possible improvements in separating the accreted component from stars formed in situ using observational proxies. Given that a large part of the mass being missed is at relatively small radii, we first consider whether targeting edge-on rather than face-on galaxies helps mitigate confusion with the disk. Second, we consider whether there is a non-parametric way to determine which regions of a surface-density map correspond to halo rather than disk stars by looking for inflection points in the cumulative distribution function (CDF) of the surface densities of the pixels. Third, we look at the distribution of metallicities in the field of view to determine whether this additional information could help determine which regions of a galaxy are more likely to be made up of accreted stars.

\subsection{Edge-on galaxies}
\label{sec:edgeon}

Our simulated galaxies pose a different but equally frustrating set of problems for estimating \fhalo\ when viewed edge-on, as illustrated in Figure \ref{fig:edge-on-sucks-too}. In most of the galaxies there is a clear delineation, especially in the direction perpendicular to the disk, between material with $d_{\mathrm{form}}>30$ kpc and $d_{\mathrm{form}}<30$ kpc. However, there is still a fairly wide variety of distributions of in situ material. In some cases like  \textsf{m12i} there is substantial in situ material at higher latitudes above the disk plane (i.e., an extended thick disk) and in a spheroid, while in others like \textsf{m12b} there is virtually no region outside of the thin disk that is dominated by stars formed in situ. In still other cases (like  \textsf{Robin}) an interaction with a massive satellite (apparent in the image) has pulled in situ material out of the disk plane, as has been recently predicted for the Sagittarius dwarf galaxy in the MW, which is of a similar mass \citep{laporte16,gomez17a,gomez17b}. While in this particular case it is easy to tell this is happening because the interacting satellite is clearly visible, such an effect is more difficult to pinpoint at later stages of such a merger, as is probably the case in \textsf{m12c}. In yet other cases (like \textsf{m11f}) there is little accreted material visible in this view at all. Therefore, although in general the in situ material appears to dominate a lower fraction of pixels in the edge-on case, the prospect of fitting a profile and choosing a threshold beyond which stars are part of the halo, as in \mer , seems to be equally problematic here. On the other hand, the dominance of the accreted component in this view validates the strategy of the GHOSTS team \citep{harmsen17} to sample along the minor axis of edge-on galaxies in order to optimize for the highest possible fraction of stars that were unambiguously formed outside the galaxy and then accreted, with the caveat that for small $z\sim R$ there is still a substantial fraction of in situ scattered material present in many of our simulated galaxies.

\begin{figure*}
\begin{tabular}{cc}
\includegraphics[width=0.45\textwidth]{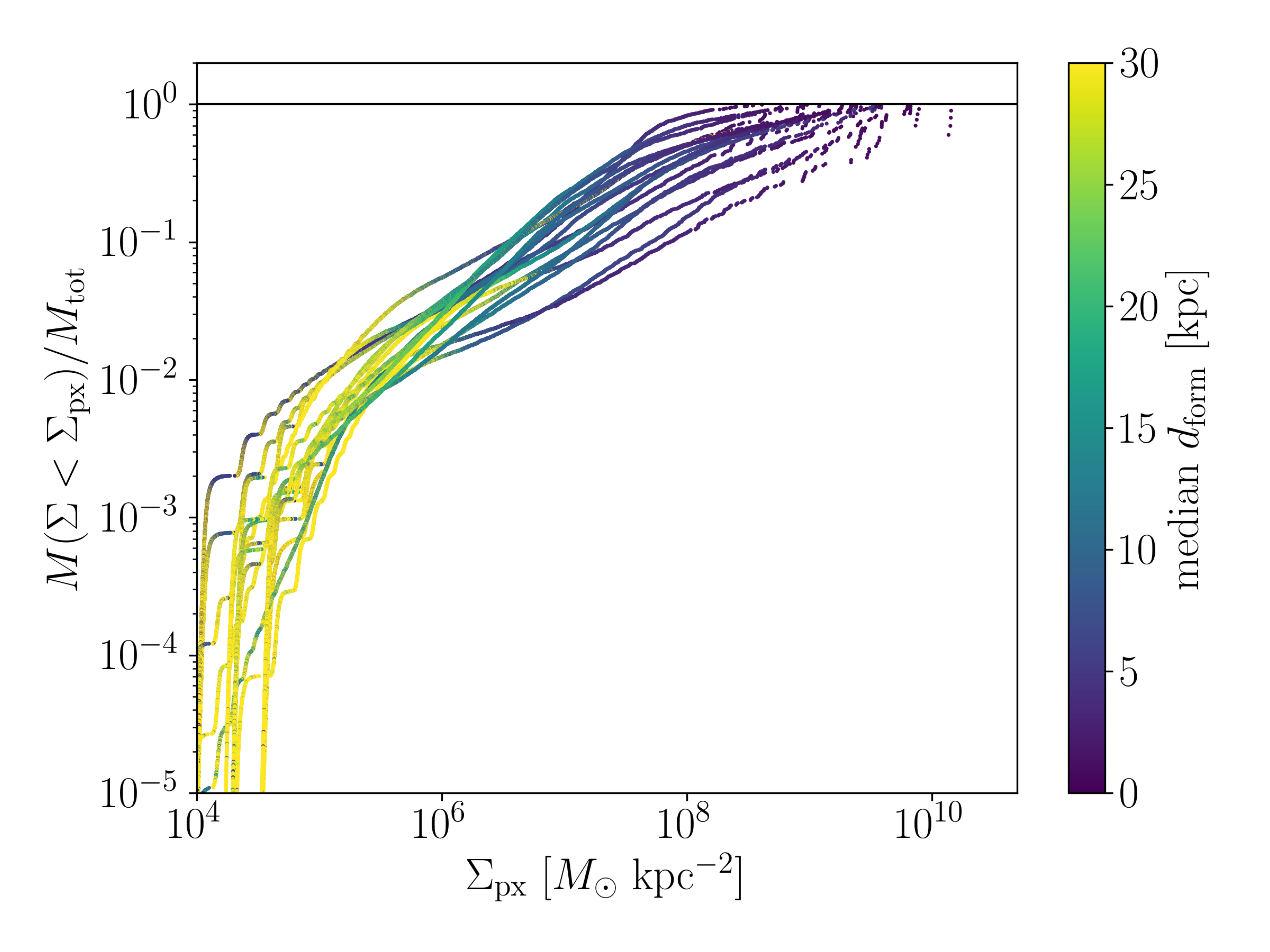} &
\includegraphics[width=0.45\textwidth]{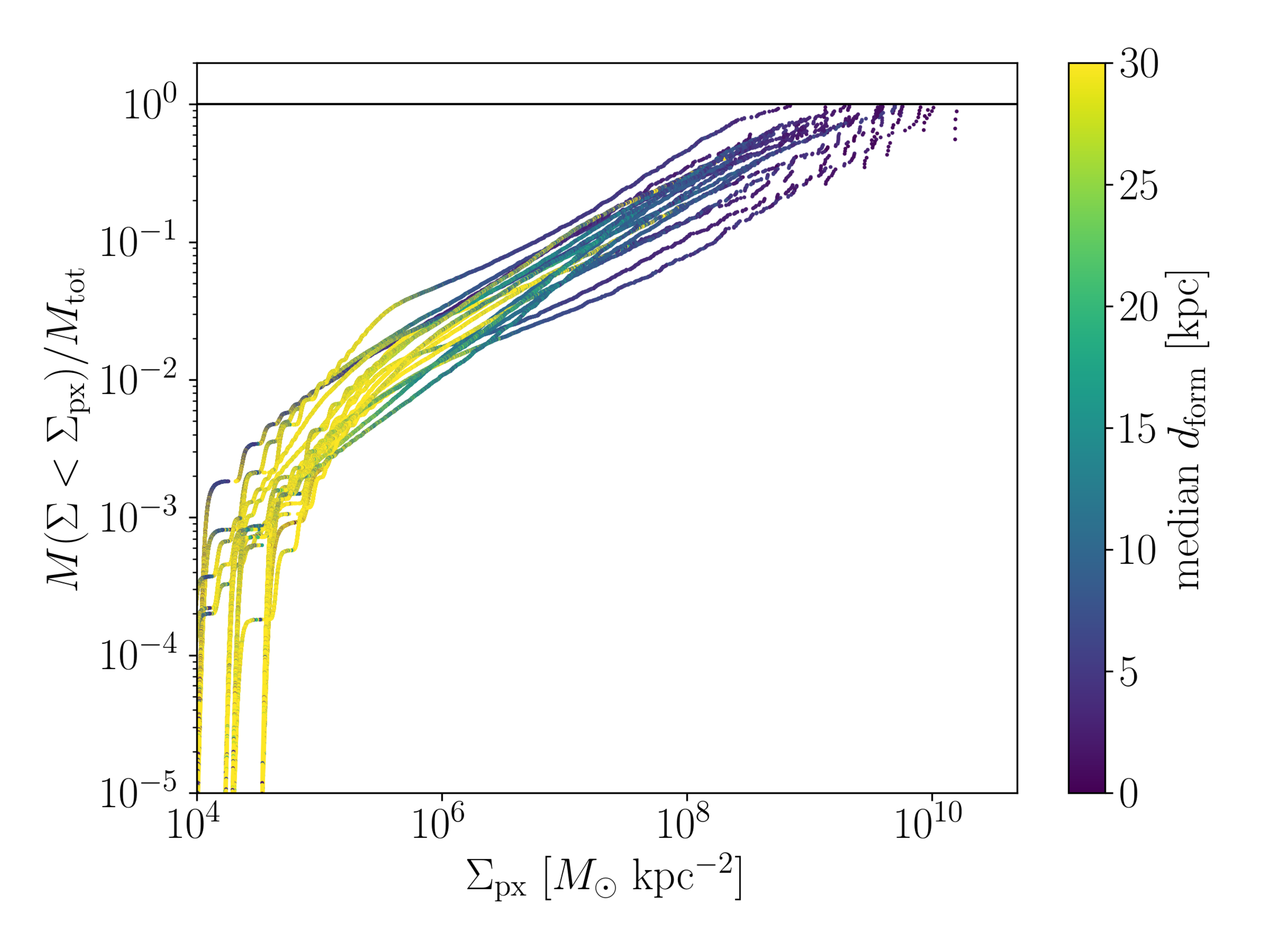}
\end{tabular}
\caption{Cumulative distribution function (CDF) of the surface mass densities in the pixels of each mock image for both face-on (left) and edge-on (right) views of the same sample of galaxies.The appearance of step features at low surface density indicates the resolution limit (pixels containing only one or a few particles). The surface density corresponding to the transition to an accreted component varies substantially from galaxy to galaxy, with no obvious transition in the CDF corresponding to a transition in the origin of the material.}
\label{fig:sigma-cdfs}
\end{figure*}

\subsection{Stellar mass density distributions}
\label{sec:cdfs}
We next examine the full distribution of stellar mass densities in the pixels of each mock image, to explore \cite[following][who studied more massive galaxies]{cooper13} whether an inflection point in this distribution will naturally differentiate between the concentrated stars in the disk and bulge and the more diffuse halo. Figure \ref{fig:sigma-cdfs} shows the cumulative distribution function (CDF) of the surface mass densities in the pixels of each mock image for both face-on (left) and edge-on (right) views. The line color is proportional to the mass-weighted average formation distance for material in each pixel. Unfortunately, no obvious criterion, such as a characteristic change in the CDF slope, appears to consistently delineate pixels dominated by accreted or in situ material: some galaxies are dominated by in situ material down to very low surface mass densities, while others are dominated by accreted material to surface densities as high as $10^7\ M_{\odot}$ \unit{kpc}{-2}. Some CDFs do appear to exhibit an inflection point around the transition between in situ and accreted material, but most do not. This is true whether the galaxies are viewed edge-on or face-on, and suggests not only that a single surface mass density cut is not suitable, but that the surface density tends to smoothly connect between pixels containing mostly accreted halo and those containing mostly in situ material, with no ``break'' or other obvious feature in the light profile or CDF.

\subsection{Metallicity}
\label{sec:metallicity}

We next consider whether spatial variations in metallicity can indicate the region where most of the stars are accreted, inspired by works like \cite{2013MNRAS.432.3391T}, \citet{ds14}, \citet{harmsen17} and \citet{ds17}. For this analysis we consider only simulations from our suite that include subgrid metal diffusion, to reduce artificial numerical noise in the distribution of [Fe/H]. This subset has similar properties to the full sample (see Appendix \ref{appx:numerics}). The range of metallicities in the halos of the different simulated galaxies is substantial, reflecting the wide variety of accretion histories in the sample: some (like \textsf{m12i}) have very well-defined, metal-poor halo regions while in others (like \textsf{m12f}) the halo is much closer in metallicity to the disk.

\begin{figure*}
\includegraphics[width=\textwidth]{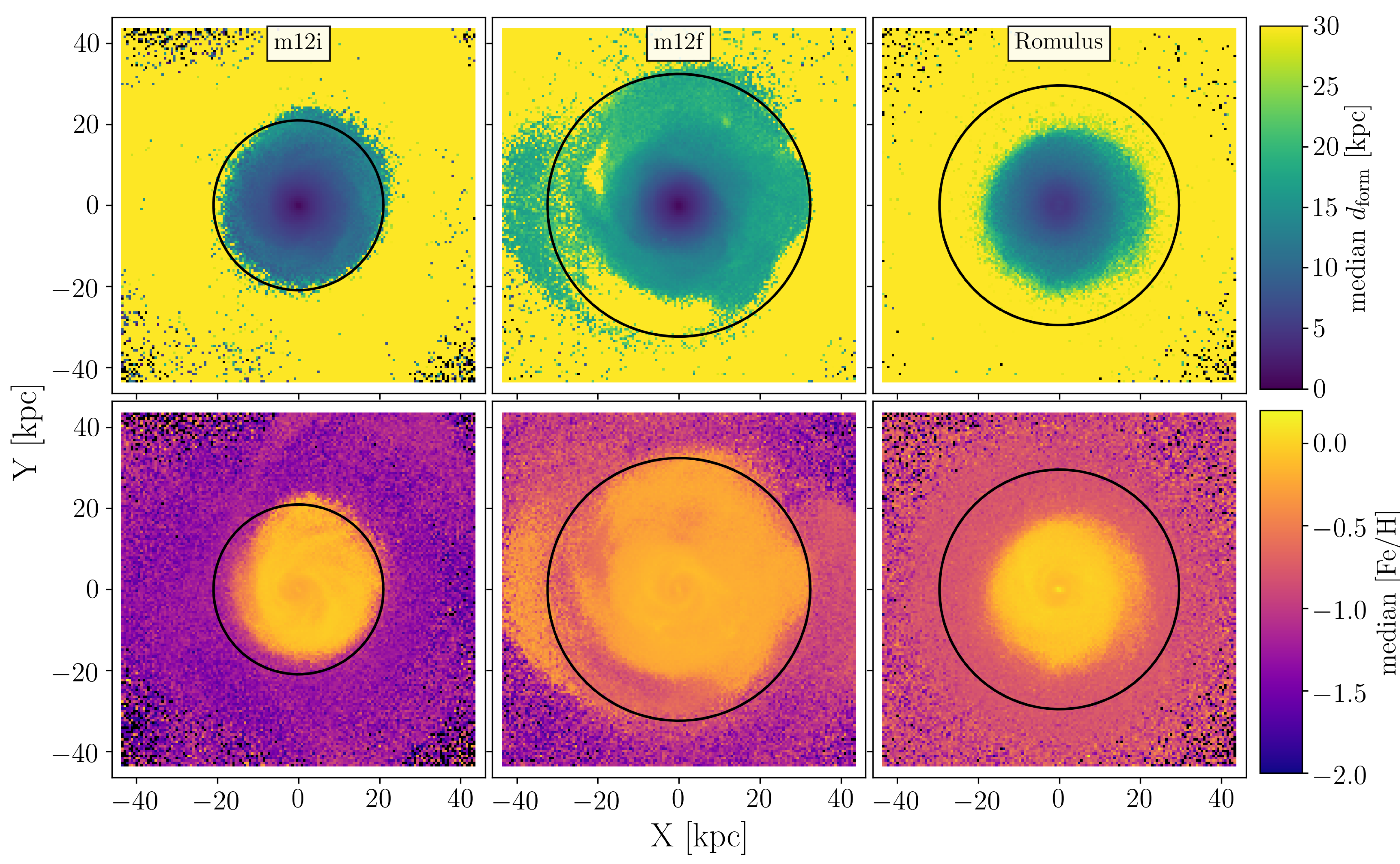}
\caption{Median formation distance (top) and median metallicity (bottom) for three simulated galaxies viewed face-on. Black pixels contain no star particles. As in Figure \ref{fig:face-on-images}, the black circles indicate $5R_d$. There is a clear contrast between accreted and in situ material apparent in the metallicity map, which suggests that it may be a useful diagnostic for distinguishing between in situ and accreted components in MW-mass galaxies.}
\label{fig:feh-avg}
\end{figure*}

Figure \ref{fig:feh-avg} suggests that the metallicity distribution can indeed be diagnostic, since in general the disk (i.e. in situ) regions have higher metallicity than the accreted outskirts regardless of their absolute metallicity, which varies substantially. Structures clearly due to mergers stand out as higher-metallicity regions on a lower-metallicity background. Information on metallicity is therefore a valuable complement in understanding which regions of a galaxy were accreted \citep[see also][]{2017ApJ...845..101B}. The obvious next step of testing whether photometric metallicities are sufficient would require modeling the SEDs of the simulated galaxies, a task we defer to future work.

\section{Conclusions}
\label{sec:conclusion}

In this paper we compared the stellar halos of high-resolution simulated galaxies from the FIRE-2 suite (\S\S \ref{sec:sims}--\ref{sec:simhalos}) with recent measurements by \citeauthor{m16} (\citeyear{m16}; \mer) of stellar halo mass fractions for eight Milky-Way-like galaxies. Because of the high resolution of these simulations, and thanks to the clear and detailed description in \mer\ of their analysis, we were able to reproduce the same steps they used to measure the stellar halo mass fractions of our simulated galaxies (\S\S \ref{sec:mockdata}--\ref{sec:fits}), and found that these had similar magnitude and scatter to the observations (\S \ref{sec:fhalo-rh}). 
 
In the simulations we can record whether each star particle was formed in situ or accreted. Thus, we can consider how well the stellar halo mass estimated from deep images agrees with the mass of accreted stars, an important theoretical quantity that is not directly accessible observationally. Inspired by \mer\ and other recent works, we considered both spatial selection based on modeling and subtracting the disk and bulge regions (\S \ref{sec:fhalo-rh}) and a selection based on surface mass density as proposed in \citet{cooper10} (\S \ref{sec:fhalo-sigma}). We found that these methods can underestimate the accreted stellar mass, usually by a factor of 2--3 and up to a factor of 10. In selections based on disk fitting and subtraction, this can introduce spurious correlations into the mass fraction that obscure the true dependence of accreted mass fraction on stellar mass. Although S\'ersic profiles can be notoriously degenerate, we found that determining the disk scale radius $R_d$ was fairly robust and not the main source of trouble. Instead, the main issues with using $R_d$ to select a region of a galaxy containing mostly accreted stars are the confusion of accreted and in situ material in the outskirts of the disk and the obscuration of accreted material by the disk at small projected radii. The contribution of each of these two complications varies substantially from galaxy to galaxy based on the individual accretion history and the morphology of the central galaxy. Using edge-on rather than face-on systems (\S \ref{sec:edgeon}) helps in the sense that the disk obscures less of the halo, but does not address the problems posed by the variety of accretion histories in selecting the accretion-dominated region. Because of the wide variety of accretion histories spanned by our set of 17 simulated galaxies, there is no characteristic surface mass density or inflection point in the distribution of stellar mass surface densities that could help systematically diagnose the presence of mostly accreted material (\S \ref{sec:cdfs}). Given the narrow mass range (less than an order of magnitude) of this set of simulations, we emphasize that this significant variation in halo properties (over three orders of magnitude in surface mass density) is \emph{not} simply an effect of mass dependence, but reflects the wide variety of accretion histories across galaxies of similar masses. If only integrated-light images are available, we suggest that examining the spatial variation in the metallicity of the galaxy, accessed perhaps through colors, is a promising approach to identifying the accreted component using minimal additional information, since the simulated galaxies show a strong contrast in metallicity between the accreted stars and those formed in situ (\S \ref{sec:metallicity}). Quantifying this relationship will be the subject of future work.

Given the limitations of any suite of simulated galaxies, these conclusions come with a few caveats. Even in our highest-resolution simulations we can only satisfactorily resolve star formation in satellite galaxies down to the mass scale of the classical dwarfs \citep[$M_{\mathrm{halo}} \gtrsim 10^9 M_{\odot}$; $M_{\mathrm{star}} \gtrsim 10^5 M_{\odot}$;][]{wetzel16},  so we are still missing some of the mass in accreted stars, but we expect this fraction to be  small given that the mass-to-light ratio of galaxies increases so sharply below this level \citep[e.g.][]{2013ApJ...762L..31B,2014ApJ...784L..14B,2014MNRAS.444..222G,2017MNRAS.464.3108G,2010ApJ...710..903M}, and given that the total stellar halo masses are $10^9 M_{\odot}$ or above for all the galaxies. On the other hand, unlike \citet{pillepich14}, we do not remove stars in bound subhalos, so this likely raises the mass fractions somewhat for a few of our simulated galaxies, notably those where a companion is clearly visible. There are not too many of these cases in our suite, which is reasonable given that the field of view shows the inner $\sim 50$ kpc where most structures will have been tidally disrupted \citep{2017MNRAS.471.1709G}. \mer\ do not attempt to fit and remove any satellites in their measurements, but neither are any companion galaxies clearly apparent in their images. The main point is to maintain consistency in the treatment of satellites between analyses of the observations and simulations, so where some of our simulated galaxies have companions we consider these part of the scatter. 

Most of our simulated galaxies have somewhat higher stellar masses than the bulk of the observed sample in \mer, a limitation shared by many of the other simulations to which those observations were compared (most of which, like many of our simulated galaxies, were originally matched to the Milky Way's properties). The most massive galaxy in the \mer sample has a stellar mass of $9\times 10^{10}$ $M_{\odot}$, which is less massive than half the simulated galaxies we consider, while their lowest-mass galaxy, at about $1.5 \times 10^{10}$ $M_{\odot}$, is only half as massive as our least massive simulated galaxy (see Figure \ref{fig:halo-fractions}). The simulated systems in the lower end of the mass range differ quite widely in terms of their present-day appearance (bulge- or disk-dominated) and their accretion histories, and clearly a larger sample than ours is needed to get a real sense of the scatter inherent at this mass scale. Work with the Illustris simulations \citep{pillepich14,2018MNRAS.479.4004E} is complementary to this study: thanks to the necessary trade-off between simulation box size and resolution, such work can more systematically explore the scatter and mass-dependence of stellar halo masses, but the star particle masses are too high to produce images, like those in Figure \ref{fig:face-on-images}, with resolved structure at the surface densities reported in \mer .

Finally, in determining whether stars were formed in situ or accreted, we used a constant cutoff in the distance from the main galaxy where each star particle was formed, as in \citet{2017ApJ...845..101B}, rather than taking a more detailed particle-tracking approach, as in previous studies \citep{font11, pillepich15, 2016MNRAS.458.2371R,aa17}. Some star particles may therefore be mistaken for accreted rather than in situ or vice versa if they came in very early and/or formed during the infall of a gas-rich galaxy. This includes only a small fraction of stars for most of our galaxies, and in some cases it is genuinely debatable what should be considered accreted or formed in situ; our method does not allow for any nuance in this area. An example in Figure \ref{fig:dform-vs-dpresent} is the infalling satellite in \textsf{Robin}, in the lower left-hand corner, where star formation clearly proceeded along with the merger. Our simplistic 30 kpc distance cutoff is clearly not the right choice in this case, but one could argue for interpreting those stars either as formed in situ (since the stars were formed when the satellite galaxy was clearly within, and influenced by, the halo of the main galaxy) or accreted (since they formed in a significantly different environment than the bulk of the stars in the main galaxy). However, given that we are comparing mainly total masses, and given the relative insensitivity of our results to the exact value of \dform that distinguishes accreted stars, we consider that this approximation is likely sufficient for the present study. In future work, we plan to extend the current analysis by using particle tracking to take into account the full time history of the star particles that end up in the outskirts of our simulated galaxies. 

Our results underline the importance of parallel analysis of observations and simulations of galaxies as the way forward in robustly comparing the two to construct tests of cosmological models of galaxy formation. In this study, such an approach revealed the difficulty in choosing a single scale or scales for apertures around galaxies that are most sensitive to a given stellar population of theoretical interest (i.e. accreted or formed in situ), at least for galaxies in the mass range of the Milky Way (and hence the targets observed in \mer). At higher mass scales this may be a more viable solution: \citet{2018MNRAS.475.3348H} explored the use of physically constant-size apertures (10 and 100 kpc) around massive elliptical galaxies in a related strategy, based on indications from simulations of elliptical galaxies in e.g. \citet{2016MNRAS.458.2371R}. However, such a method is unlikely to be viable for galaxies at lower mass scales, where morphologies and sizes differ quite widely: as seen in Table \ref{tbl:sims}, the radius enclosing 90 percent of the stellar mass in our sample varies over an order of magnitude even for our relatively narrow mass range due in part to the variety of formation histories \citep{2017arXiv171203966G}, so choosing a single set of physical apertures is likely to yield even worse results than scaling the aperture to a fitted disk scale length. Simultaneous analysis of real and simulated observations of stellar halos can thus also indicate which proxies are most appropriate for separating accreted stars from those formed in situ at different mass scales, since galaxies certainly vary widely with stellar mass. These valuable insights, enabled by the match between state-of-the-art observation and simulation techniques, are best gained by the type of in-depth conversation between simulators and observers that inspired this work.

\section*{Acknowledgements}
The authors thank Laura Sales, Lydia Elias and Kyle Stewart for helpful suggestions during the GalFRESCA 2017 meeting organized by Shea Garrison-Kimmel and Coral Wheeler, and Alison Merritt for helpful background on her measurement techniques. 

RES was supported by an NSF Astronomy \& Astrophysics Postdoctoral Fellowship under grant AST-1400989. 

Support for SGK was provided by NASA through Einstein Postdoctoral Fellowship grant number PF5-160136 awarded by the Chandra X-ray Center, which is operated by the Smithsonian Astrophysical Observatory for NASA under contract NAS8-03060.

AW was supported by a Caltech-Carnegie Fellowship, in part through the Moore Center for Theoretical Cosmology and Physics at Caltech, and by NASA through grants HST-GO-14734 and HST-AR-15057 from STScI. 

TKC was supported by NSF grant AST-1412153.

Support for PFH was provided by an Alfred P. Sloan Research Fellowship, NSF Collaborative Research Grant \#1715847 and CAREER grant \#1455342. 

DK acknowledges support from NSF grants AST-1412153 and AST-1715101 and the Cottrell Scholar Award from the Research Corporation for Science Advancement.

IE was supported by a Ford Foundation Predoctoral Fellowship.

CAFG was supported by NSF through grants AST-1412836, AST-1517491, AST-1715216, and CAREER award AST-1652522, and by NASA through grant NNX15AB22G.

Numerical calculations were run on the Caltech compute cluster ``Wheeler,'' allocations from XSEDE TG-AST130039 and PRAC NSF.1713353 supported by the NSF, NASA HEC SMD-16-7592, and the High Performance Computing at Los Alamos National Labs.



\bibliographystyle{mnras}
\bibliography{references}
 

\appendix

\section{Numerical effects}
\label{appx:numerics}
\begin{figure}
\begin{center}
\includegraphics[width=0.5\textwidth]{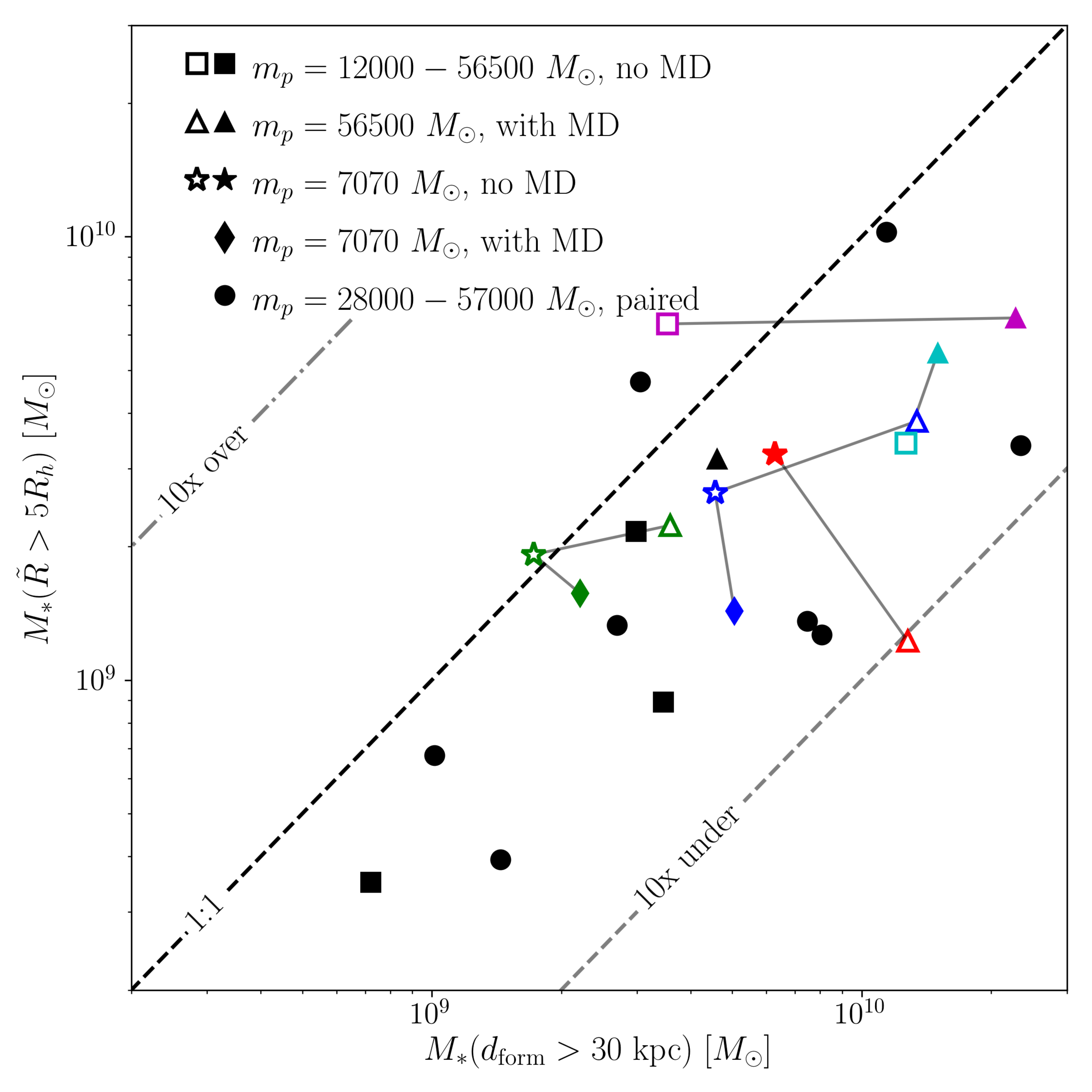}
\caption{Estimated mass (as in the right-hand panel of Figure \ref{fig:mass-estimates}, but using the $\tilde{R}>5R_h$ criterion only) relative to the ``true'' accreted stellar mass (formed beyond 30 kpc) for the simulated galaxies in Table \ref{tbl:sims} (filled symbols) and companion simulations with the same initial conditions but different resolution and/or physics (open symbols). Points in the same color have the same initial conditions, and are connected by grey lines. ``MD'' in the caption indicates whether a numerical implementation of subgrid turbulent metal diffusion is included in the simulations. Different resolutions and physics produce no significant trends in the relative accuracy of estimates of the accreted stellar mass using this criterion.}
\label{fig:sim-masses}
\end{center}
\end{figure}

Figure \ref{fig:sim-masses}, similar to the right-hand panel of Figure \ref{fig:mass-estimates}, shows the estimated mass using the $\tilde{R}>5R_h$ criterion, relative to the mass formed beyond 30 kpc for different resimulations of the galaxies in Table \ref{tbl:sims} that use different resolution and/or subgrid physics. The filled symbols are the set used in the main paper and listed in the table. As was pointed out in \cite{hopkins17}, higher-resolution simulations (stars and diamonds) tend to have slightly lower total stellar masses than lower-resolution versions (squares and triangles) with the same initial conditions, and this also appears to be true for the mass in accreted stars. However this trend is quite noisy and probably complicated by differences in the treatment of turbulent metal diffusion between runs at different resolution. To illustrate, we consider the two sets of simulations shown in the figure, \textsf{m12i} and \textsf{m12f}, that have been run with resolutions lower by a factor of 8 relative to the fiducial run, with an otherwise identical setup, permitting a controlled comparison. In one case, lowering the resolution by this factor increases the mass formed beyond 30 kpc by 60 percent, while in the other, it is lower by 40 percent relative to the fiducial run used in our analysis. Resolution does not appear to significantly affect the degree to which the selection criterion over- or under-estimates the mass in accreted stars relative to its true value. 

Simulations including subgrid turbulent metal diffusion (triangles and diamonds) look to perhaps have slightly \emph{higher} accreted masses than their counterparts without this additional physics (squares and stars). Comparing different versions of \textsf{m12i} and \textsf{m12f} again, but now selecting those that have been run at the same (highest) resolution while varying only the treatment of metal diffusion, the mass formed beyond 30 kpc is 22 and 10 percent lower in the runs without metal diffusion relative to fiducial, respectively. Varying the diffusion prescription likewise shows no trend in terms of estimator accuracy, so we can safely assume that the results of the subset discussed in \S \ref{sec:metallicity} extend to our entire sample. 

Finally, we find that paired and isolated halos are also similarly distributed within the scatter (which is again likely to be partially attributable to the variation in resolution and metal diffusion treatments, as discussed above). Based on these findings, we do not expect that our results are highly affected by numerics.

\section{Monte-Carlo S\'ersic fit test}
\label{appx:fit-test}
\begin{figure*}
\begin{center}
\includegraphics[width=\textwidth]{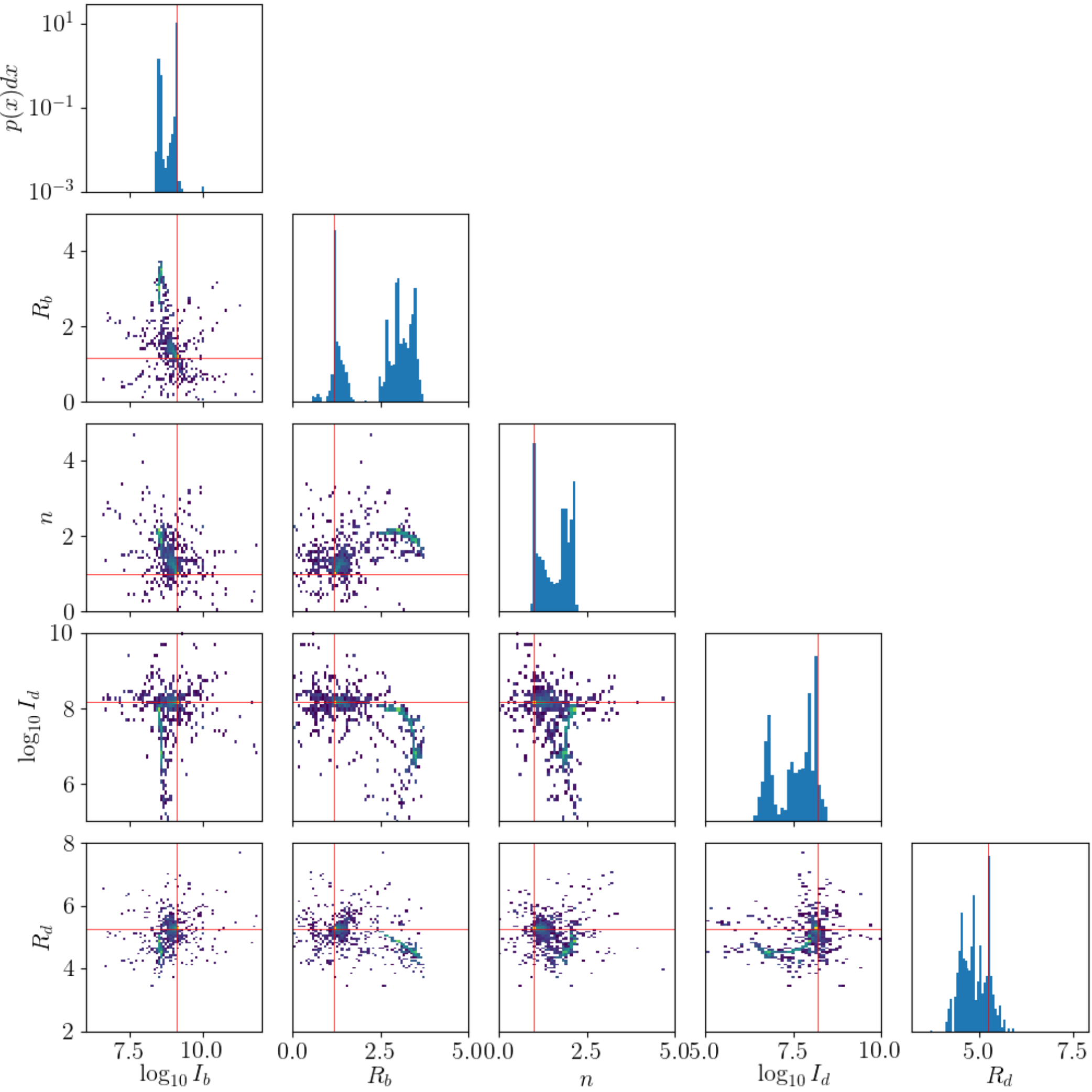}
\caption{MCMC exploration of the five-dimensional parameter space of the double S\'ersic profile for the simulated galaxy \textsf{m12i}. Details of the sampler setup are given in \S \ref{sec:fits}. The two-dimensional projections show the binned samples on a logarithmic color scale from low (blue) to high (yellow) density. The histograms show one-dimensional projections on a logarithmic scale with identical limits on the y axis. The median of all the samples along each parameter is marked by red lines.  The distribution of samples is multivalued in most projections, with the most probable of the peaks also displaying the least degeneracy. }
\label{fig:mcmc-results-large}
\end{center}
\end{figure*}

In order to understand how degeneracies between the 5 S\'ersic parameters could affect the fit results, we carried out a more complete 5-dimensional exploration of parameter space using the affine-invariant sampler \textsf{emcee}, starting from a broad distribution in a ball around the best fit reported by the local Levenberg-Marquardt minimization. We used 50 walkers to sample the parameter space 10000 times per walker, discarding the first 200 steps per walker as burn-in. The distribution of the remaining samples is shown in Figure \ref{fig:mcmc-results-large}. There are often multiple peaks in most projections of the parameter space, but one peak (at the intersection of the red lines) is much higher and narrower than the other and displays less degeneracy between parameters. It is also clear that the more-degenerate region spans to much less sensible values of some of the parameters, notably the disk normalization $I_d$ and the bulge scale radius $R_b$. The normalization of the disk profile is of special concern since it could potentially bias the disk scale radius $R_d$, but in fact this parameter is relatively robust over a broad range of values, which gives us some confidence that a more localized minimization algorithm will not obtain wildly different values for $R_d$ depending on which local minimum it ends up in, even if the other parameters are more strongly affected.

The taller peak is extremely narrow and not very degenerate between different parameters except for in the extreme tails of the $I_d$--$R_d$ projection. The narrowness is probably due in part to the lack of a proper treatment of errors in the least-squares likelihood, where we simply use the square root of the mass surface density. However the well-confined nature of this peak shows that it is a robust solution to minimizing least-squares differences, and its relative prominence and symmetrical shape suggests that one can tell when a fit by a local minimization algorithm has obtained it by starting with a good guess, considering the results of some perturbations to the initial location of the best fit, and making sure that the normalizations to the two components have reasonable best-fit values. Additionally, the most important parameter for this analysis, the disk scale length $R_d$, is not very susceptible to variations in the other best-fit parameters. We therefore use the results from the Levenberg-Marquardt minimization, shown in Table \ref{tbl:fit-results}, with confidence.


\end{document}